\documentclass[a4paper,11pt]{article}
\pdfoutput=1 

\usepackage{jheppub} 
\usepackage{amsmath,amssymb,graphicx,slashed,xcolor}
\usepackage{dsfont}
\usepackage{stackengine}
\usepackage{accents}

\usepackage[T1]{fontenc} 

\newcommand{\refeq}[1]{(\ref{#1})}

\numberwithin{equation}{section}
\newcommand{\nn}{\nonumber \\}
\newcommand{\beq}{\begin{equation}}
\newcommand{\eeq}{\end{equation}}
\newcommand{\ba}{\begin{array}}
\newcommand{\ea}{\end{array}}
\newcommand{\bea}{\begin{eqnarray}}
\newcommand{\eea}{\end{eqnarray} }
\newcommand{\bal}{\begin{align}}
\newcommand{\eal}{\end{align}}

\newcommand{\Lcal}{\mathcal{L}}

\newcommand{\Ocal}{\mathcal{O}}
\newcommand{\Dcal}{\mathcal{D}}

\newcommand{\Ccal}{\mathcal{C}}
\newcommand{\Mcal}{\mathcal{M}}

\newcommand{\Wcal}{\mathcal{W}}

\renewcommand{\d}{\partial}

\title{\boldmath Wilsonian renormalisation of CFT correlation functions: Field theory}

\author{J. M.\ Lizana}
\author{and M.\ P\'erez-Victoria}

\affiliation{CAFPE and Departamento de F\'{\i}sica Te\'orica y del Cosmos,  Universidad de Granada, Campus de Fuentenueva, E-18071, Granada, Spain}

\emailAdd{jlizan@ugr.es}
\emailAdd{mpv@ugr.es}

\abstract{
We examine the precise connection between the exact renormalisation group with local couplings and the renormalisation of correlation functions of composite operators in scale-invariant theories. A geometric description of theory space allows us to select convenient non-linear parametrisations that serve different purposes. First, we identify {\em normal} parameters in which the renormalisation group flows take their simplest form; normal correlators are defined by functional differentiation with respect to these parameters. The renormalised correlation functions are given by the continuum limit of correlators associated to a cutoff-dependent parametrisation, which can be related to the renormalisation group flows. The necessary linear and non-linear counterterms in any arbitrary parametrisation arise in a natural way from a change of coordinates. We show that, in a class of minimal subtraction schemes, the renormalised correlators are exactly equal to normal correlators evaluated at a finite cutoff. To illustrate the formalism and the main results, we compare standard diagrammatic calculations in a scalar free-field theory with the structure of the perturbative solutions to the Polchinski equation close to the Gaussian fixed point.
}

\begin{document} 
\maketitle
\flushbottom

\section{Introduction}
\label{sec:i} 

The process of renormalisation in quantum field theory is intimately connected with the concept of renormalisation group (RG) invariance. From the Wilsonian point of view, renormalisable theories can be understood as sets of points in theory space that are reached by exact RG transformations from the neighbourhood of a particular ultraviolet (UV) fixed point. Furthermore, because the continuum limit can be written as an infinite RG flow towards the infrared (IR), the cutoff dependence of the bare couplings necessary to renormalise the theory is essentially determined by the form of the RG flows near the fixed point. One important consequence of this is the original formulation of the RG for renormalisable theories, in which a change of renormalisation scheme (in particular a change of renormalisation scale) can be compensated by a change in a finite number of renormalised couplings, in such a way that the observables are kept invariant. 

Even if this qualitative Wilsonian view of renormalisation is nowadays widely appreciated, some fundamental issues have not yet been worked out in full detail. The main question we want to address in this paper is the following: i) what are the implications of the exact RG on the renormalisation of correlation functions of arbitrary composite operators? And conversely, ii) what can we learn about the exact RG from the structure of the counterterms that renormalise these operators? 

A systematic functional formalism to renormalise general correlators of composite-operator correlators was developed some time ago by Shore in~\cite{Shore:1990wp} (see also~\cite{Osborn:1989td,Jack:1990eb} for previous developments). Let us summarise this approach. First, a source is introduced for each of the (infinitely-many) independent local operators in the theory~\cite{Drummond:1977dg}. This is equivalent to a local version of theory space in which the couplings of the different operators can depend on the space-time point. Renormalisation then proceeds quite standardly by writing the bare (space-time dependent) couplings as convenient functions of renormalised (space-time dependent) couplings. At the linear level, this is equivalent to the usual multiplicative renormalisation of the operators, including mixing. Further counterterms are required to make finite the correlators of two or more composite operators. The main point of~\cite{Shore:1990wp} is that all these counterterms can be generated by a non-linear dependence of the bare couplings on the renormalised ones. In general, the bare couplings at a point $x$ depend not only on the values of the renormalised couplings at $x$ but also on its derivatives at this point. Because we want to compare with the exact RG, we will use a dimensionful regulator. Then, mixing of operators of different dimension is expected already at the linear level. 

In order to understand these features of the renormalisation of composite operators from the point of view of the exact RG, it is clear that we need to consider RG flows in a theory space with space-time dependent couplings.\footnote{This should be distinguished from the local RG~\cite{Osborn:1991gm}, which goes one step further and studies evolution under Weyl transformations. We will restrict our attention to the usual RG evolution under global dilatations.}  As we will see in detail, such local RG flows can be used to define space-time dependent bare couplings that renormalise the theory. Both the linear and the non-linear terms in the renormalised couplings follow automatically. This relation had been found before in~\cite{Polonyi:2000fv}, but only at the linear level and without taking into account the effect of derivatives in the evolution, which turns out to be crucial. Our analysis will have no restrictions in this sense. On the other hand, to simplify the discussion and the equations, in this paper we focus on correlation functions at a fixed point of the RG. 

Fixed points correspond to scale invariant theories, which, if unitary, are also expected to be conformally invariant~\cite{Polchinski:1987dy,Dymarsky:2013pqa,Nakayama:2013is}. A lot is known about correlation functions in conformal field theories (CFT). In particular, there has been recent progress in the bootstrap program, which aims to determine the correlation functions from minimal input and consistency conditions~\cite{Ferrara:1973yt,Polyakov:1974gs,Mack:1975jr,Belavin:1984vu,Dolan:2003hv} (see also~\cite{Rychkov:2016iqz} for references to more recent work). These methods refer only to the fixed-point itself, whereas in this paper we are investigating the relation with finite deformations in the presence of a cutoff. Of course, after renormalisation the results for infinitesimal deformations must agree, so the CFT consistency conditions will impose non-trivial constraints on the behaviour of the RG flows near the fixed point.

When the fixed point of interest is Gaussian, the renormalisation procedure takes its simplest form when formulated as a limit of deformations of the fixed point. Then the connection with the RG is most transparent. However, for interacting CFT it may be more convenient to consider deformations of another separate point in theory space, if they are described by simpler actions. This point must belong to the basin of attraction of the fixed point, i.e.\ to the so-called critical manifold. In the paper we consider this possibility. It will be crucial for the application of these methods to holographic renormalisation in gauge/gravity duality. This application is actually the main motivation for this work and will be presented in a companion paper~\cite{LPVholography}.

We will formulate both the renormalisation process and the exact RG in a geometric language, very similar to the one developed in~\cite{Dolan:1994pq} (see also \cite{Knizhnik:1984nr,Kutasov:1988xb,Sonoda:1992hd,Ranganathan:1992nb,Ranganathan:1993vj}).\footnote{The main difference in our formalism is that we incorporate the space-time dependence of the couplings into the geometry of theory space, which allows for general quasilocal changes of coordinates. Furthermore, we work with a dimensionful cutoff regularisation, which is also included in the description of theory space.} In this formulation, the space-time dependent couplings are understood as coordinates of theory space, the beta functions are vector fields, the operators are vectors and the correlation functions are tensors. Special attention will be paid to the active role of the cutoff in the parametrisation of theory space. Writing the equations in a coordinate-invariant fashion will allow us to easily change coordinates to find parametrisations that suit different purposes.  In particular, we define the renormalised couplings as a set of cutoff-dependent coordinates. Then the renormalised correlators can be interpreted as correlators in this ``renormalisation'' coordinate system. These special coordinates will be used to define a flat connection that contains all the information about the non-linear counterterms that cancel coincident-point divergences in arbitrary parametrisations. 

Changes of coordinates can also be useful to put the exact RG flows in a manageable form. In fact, we shall identify {\em normal coordinates} around a given fixed point, in which the beta functions and RG flows are particularly simple. At the linear level, this reduces to identifying the deformations of the fixed point that are eigenfunctions of the linearised RG evolution. These deformations are regularised versions of the primary operators at the fixed point, with eigenvalues simply related to their scaling dimensions.\footnote{We also allow for the possibility of non-diagonalisable linear terms, which give rise to logarithmic CFT.}  The normal coordinates are an extension of this linear behaviour. When the dimensions take generic values, they are such that all the non-linear terms in the flows vanish. For exceptional values of the dimensions, on the other hand, non-linear terms are unavoidable but can be reduced to a minimal set. These terms give rise to the usual Gell-Mann-Low beta functions of mass-independent schemes and to conformal anomalies.

As a consequence of the intimate connection with the exact RG, the counterterms can be found from the RG flows. Furthermore, the renormalised correlators turn out to be equal to specific bare correlators evaluated at a finite cutoff. Of course, the determination of counterterms and renormalised correlators from RG flows cannot be unique, since there is some freedom in the renormalisation process. Different choices give rise to different renormalisation schemes and some scheme dependence survives the continuum limit and leaks into the renormalised correlators. We will pay special attention to these ambiguities and show that when they are fixed by a minimal subtraction condition, the resulting renormalised correlators are equal to bare correlators defined by functional differentiation with respect to couplings in normal coordinates, at a finite cutoff that is identified with the renormalisation scale.

The paper is organized as follows. In Sections~\ref{sec:theory} to~\ref{sec:Correlation}, we develop the formalism of Wilsonian renormalisation. In Section~\ref{sec:theory}, we define theory space in terms of actions and cutoff scales and introduce its parametrisations with local couplings. In Section~\ref{sec:RG}, we write the exact RG flows and beta functions in a geometric form and in arbitrary coordinates; we also describe their transformation under a change of coordinates. In Section~\ref{sec:normal}, normal coordinates are defined. In Section~\ref{sec:Correlation} we describe in geometric terms the renormalisation of correlators of composite operators and we make explicit its relation with the RG flows. In Section~\ref{sec:free}, we illustrate the method with a simple explicit example: we consider the theory space of a real scalar field and study the Polchinski RG flows in the neighbourhood of the Gaussian fixed point and their relation with the standard renormalisation of free-field correlators of composite operators.



\vspace{.8cm} \noindent {\sc Index notation}

\vspace{.2cm} \noindent Before getting started, let us explain a notation we employ intensively throughout this paper. Lorentz indices are as usual written with greek letters $\mu,\nu,\ldots$ We use $x,y,\ldots$ to represent $d$-dimensional space-time coordinates, which will be written sometimes as a continuous index: $g^x \equiv g(x)$. The discrete indices $a,b,\ldots$  are used to represent flavour indices (including Lorentz indices if necessary), i.e.\ they label the different operators or couplings of the theory (or their components). We also use the DeWitt condensed notation, with the index $\alpha$ (or $\alpha_i$) indicating a set of flavour and space-time indices; for instance $g^\alpha=g^{ax}=g^a(x)$. The Einstein summation convention is used for both discrete and continuous indices, with repeated space-time indices indicating an integration in that variable. As an example,
\begin{align}
k_{\alpha_1 \alpha_2} g^{\alpha_1}g^{\alpha_2} & = k_{a_1 x_1 a_2 x_2} g^{a_1 x_1} g^{a_2 x_2}  \nn
& = \sum_{a_1,a_2} \int d^dx_1 d^dx_2 |\gamma| k_{a_1a_2}(x_1,x_2) g^{a_1}(x_1)g^{a_2}(x_2) . \label{notationexample}
\end{align}
Here, $\gamma$ is the metric in the $d$-dimensional spaces parametrised by $x_1$ and $x_2$, and we have written the double $d$-form $k_{a_1 x_1 a_2 x_2}$ in terms of a tensor $k_{a_1a_2}(x_1,x_2) = k_{a_1 a_2}^{x_1,x_2}$. The square root of $|\gamma|$, the determinant of the metric, is used to raise space-time indices. Sometimes we will find it convenient, for clarity, to write integrals and the metric explicitly, as in the second line of~\refeq{notationexample}. Finally, indices inside a parenthesis label the entries of diagonal (generalised) matrices. Therefore, there is no sum or integral in the equation 
\begin{equation}
q^\alpha  = \lambda_{(\alpha)} g^\alpha,
\end{equation}
while
\beq
\lambda_{(\alpha)} k_\alpha g^\alpha = \sum_a \int d^d x \sqrt{|\gamma|} \lambda_{(a)}(x)  k_a(x) g^a(x).
\eeq 


\section{Wilson actions and theory space}
\label{sec:theory}

Consider a generic local quantum field theory in $d$ flat Euclidean dimensions, defined by a classical Wilson action $s$, and the corresponding partition function evaluated with a UV cutoff. The Wilson action is a quasi-local functional of a set of quantum fields $\omega$, 
\beq
s[\omega] = \int d^d x\, \Lcal \left(x;\omega(x),\d \omega(x),\ldots \right). \label{action}
\eeq
We have allowed for an explicit space-time dependence, which will be useful for the definition and calculation of correlation functions. The cutoff partition function is obtained by functional integration over the fields $\omega$, 
\beq
Z_\Lambda(s) = \int \left[\Dcal \omega \right]^\Lambda \, e^{-s[\omega]} . \label{partitionfunction}
\eeq
For the moment we do not need to know the nature of the regularisation; we just assume that it is characterised by the indicated cutoff scale $\Lambda$.  Let $\mathcal{I}$ be the set of all Wilson actions with field content $\omega$ and given symmetry restrictions. The theory space we will work on is given by $\Wcal = \mathcal{I} \times \mathbb{R}^+$.\footnote{This idea is analogous to working on a theory space of extended actions that implement the cutoff.} We will, somewhat loosely, treat these spaces as infinite-dimensional smooth manifolds. A point in $\Wcal$, i.e.\ an action $s$ and a scale $\Lambda$, specifies a particular theory described by $Z(s,\Lambda)\equiv Z_\Lambda(s)$. This definition agrees with the following general rule: given any map $U:\Wcal\to X$, with $X$ any set, we define $U_{\Lambda}:\mathcal{I}\to X$ by $U_{\Lambda}(s)=U(s,\Lambda)$.

There are however some redundancies in this description of the space of theories. 
In particular, 
a rescaling $x=t x^\prime$  defines the new action 
\beq
s_t[\omega] = s[D_{t^{-1}} \omega],
\eeq
where $D_t$ is a dilatation.\footnote{Remember that $\omega$ represents a set of fields, which may be scalars, tensors or spinors. Under the dilatation, which is a particular change of coordinates, each of these fields transforms in a definite way. For a tensor with $n^u$ ($n^d$) contravariant (covariant) indices, $(D_t \omega)(x)=t^{n^d-n^u} \omega(t x)$.}
Changing variables $\omega \to D_t \omega$ in the path integral and neglecting the trivial Jacobian we get
\beq
Z_\Lambda(s)=Z_{t \Lambda} (s_t). \label{equivalence}
\eeq
This defines the equivalence relation $(s_t,t \Lambda) \sim (s_{t^\prime},t^\prime \Lambda)$. It is very convenient to introduce the rescaled flat metric $\gamma^t_{\mu\nu}=t^2 \eta_{\mu\nu}$. The equivalence relation can then be understood as $(s_t,\gamma^{t}) \sim  (s_{t^\prime},\gamma^{t^{\prime}})$, where the cutoff in the partition function is to be measured in energy units associated to the metric $\gamma^t$ in the second entry: $\d^2 / (t^2 \Lambda^2) = (\gamma^t)^{\mu\nu}\d_\mu\d_\nu / \Lambda^2$.\footnote{Expressed in this form, we see that this is a particular case of a larger redundancy in general curved space-time. Given a change of space-time coordinates $x= \xi(x^\prime)$ and defining $s_\xi[\omega] = s[\omega\circ \xi^{-1}]$, we have $(s, \gamma) \sim (s_\xi, \gamma^\xi)$, with the cutoff evaluated with the indicated metrics and $\gamma^\xi_{\mu\nu} = \d_\mu \xi^\tau \d_\nu \xi^\sigma \gamma_{\tau\sigma}$. In this paper we neglect the backreaction of the fields on the metric and work only with the flat metrics $\gamma^t$.}  For some purposes it is useful to work with the quotient space $\Mcal = \Wcal / \! \sim$.  As in any quotient space, there is a projection $[\;]$ into equivalence classes: given $(s,\Lambda)\in \Wcal$,   $[(s,\Lambda)] \in \Mcal$ is the equivalence class it belongs to. Conversely, given a positive number $\Lambda$, we define $\rho_\Lambda: \Mcal \to \Wcal$ by $\rho_\Lambda (\mathbf{s}) = (s,\Lambda)$ with $[(s,\Lambda)]=\mathbf{s}$. In particular, using $\rho_1$ amounts to working with dimensionless space-time coordinates, as done in~\cite{Lizana:2015hqb}.
The partition function acting on equivalence classes is
\beq
\mathbf{Z}(\mathbf{s}) = Z \circ \rho_\Lambda(\mathbf{s}) .
\eeq

We will also be interested in the tangent bundles $\mathrm{T}\Wcal$ and $\mathrm{T}\Mcal$. Any vector $v$ in the tangent space of a given point $(s,\Lambda)\in \Wcal$ can be associated to an operator $\left.\Ocal\right|_{(s,\Lambda)}$ built with the quantum fields $\omega$. Let $S_\omega$ be the function on $\Wcal$ given by 
\beq
S_\omega (s)=s[\omega].
\eeq 
Then,
\beq
\left.\Ocal\right|_{(s,\Lambda)}[\omega] = \left. v \right|_{(s,\Lambda)} S_\omega . \label{defop}
\eeq
Note that only the vector components along the $\mathcal{I}$ directions enter in this equation.
The operator $\left.\Ocal\right|_{(s,\Lambda)}$, which could be non-local, represents an infinitesimal deformation of the action $s$. Conversely, given an operator $\Ocal[\omega]$, we can define a curve $(s+ t \Ocal,\Lambda)$ (with the natural definition of the sum of functionals) and associate the vector tangent to it at $t=0$: given a function $F$ in $\Wcal$,
\beq
\left. v \right|_{(s,\Lambda)} F = \left. \frac{\d}{\d t} F(s+ t\Ocal,\Lambda) \right|_{0}. \label{defvec}
\eeq
The relations \refeq{defop} and~\refeq{defvec} are inverse to each other if the vector $v$ is restricted to be orthogonal to the $\Lambda$ direction. So, we can use the same name for an operator and the vector along $\mathcal{I}$ identified with it, and will sometimes follow this convention.
We define in a similar way the expectation value of a functional or operator $G$ at the point $(s,\Lambda)$:
\begin{align}
\langle G \rangle_{(s,\Lambda)} &= \frac{1}{Z(s,\Lambda)}\left. \frac{\d}{\d t} Z(s- tG,\Lambda) \right|_{0} \nn
& = -\frac{1}{Z(s,\Lambda)}\left.v_{G}\right|_{(s,\Lambda)} Z.
\end{align} 
In the second line we have used the vector $v_G$, associated to $G$ by~\refeq{defvec}.

To parametrise the spaces $\Wcal$ and $\Mcal$, we use an infinite set $\Ccal$ of smooth functions $g^a:\mathbb{R}^d \to \mathbb{R}$, which can be regarded as background fields or space-time dependent couplings. Most importantly for our purposes, they can act as sources to define and calculate correlation functions. We define a class of parametrisations or coordinate systems in the following way.  We choose a quasilocal functional $S$ of fields and couplings such that, for each point $(s,\Lambda)\in \Wcal$,
\begin{align}
s[\omega] & = S[\gamma^\Lambda;g,\omega] \nn
& = \int d^d x\sqrt{|\gamma^\Lambda|} \Lcal \left(\gamma^\Lambda;g(x),\omega(x),\d \omega(x),\ldots \right), ~\label{parametrisation}
\end{align}
for some unique $g\in \Ccal$. The dimensionful metric $\gamma^\Lambda$ allows to work with couplings and fields of mass dimension $n^d-n^u$, with $n^d$ ($n^u$) the number of covariant (contravariant) indices they have. This metric and its inverse is used to contract the Lorentz indices, including those in derivatives. For instance, the standard linear parametrisation is given by
\beq
S[\gamma;g,\omega]   = \int d^d x  \sqrt{|\gamma|} g^a(x) \Ocal_a[\gamma;\omega](x) 
. \label{linearaction}
\eeq
Here, $\{ \Ocal_a  \}$ is a complete set of linearly-independent Lorentz-covariant local operators made out of the relevant quantum fields $\omega$ and their derivatives, modulo total derivatives (we do not include total derivatives of operators in this set because they can be absorbed after integration by parts into the space-time dependent couplings). Further symmetry and consistency restrictions may apply.
In this paper we mostly concentrate on Lorentz scalar operators, but we should keep in mind that this set is not stable under RG evolution. Among these operators, we include the identity operator, which contributes to the vacuum energy. We label this operator and its constant coupling with the index $a=0$.

\refeq{parametrisation} defines a (generalised) coordinate chart\footnote{For simplicity, we assume the regions of the spaces we work with can be covered by a single chart and neglect global issues throughout the paper.} $c:\Wcal \to \Ccal \times \mathbb{R}^+$, $c(s,\Lambda)=(g,\Lambda)$. We will use indices $\tilde{\alpha}$ to refer to either the label $\alpha$ in $\Ccal$ or to the $\mathbb{R}^+$ component, which we represent with the symbol $\wedge$. So, $c_{\Lambda}^{\alpha}(s)=c^\alpha(s,\Lambda)=g^{\alpha}$ and $c^\wedge(s,\Lambda)=\Lambda$. 

To simplify the formulas in the paper, we make the following definitions. First, we introduce the canonical projection $\pi: \Ccal\times \mathbb{R}^+ \to \Ccal$, $\pi(g,\Lambda)=g$ and call $c^{\pi}\equiv \pi \circ c$ and $c^{\pi}_{\Lambda}\equiv \pi \circ c_{\Lambda}$.
Second, we define the function $\bar \gamma :\Wcal \to T^0_2\left(\mathbb{R}^d\right)$,  $(s,\Lambda) \mapsto \Lambda^2 \eta_{\mu \nu} dx^{\mu} \otimes dx^{\nu}$. In terms of it, 
$2\bar\gamma\frac{\partial}{\partial \bar\gamma} F(s,\Lambda) = \Lambda \partial_{\Lambda} F(s,\Lambda)$
for any $F: \Wcal \to \mathbb{R}$.  Sometimes we will keep the coordinates $c$ implicit. In particular, we define
\beq
H^{\tilde{\alpha}} = c^{\tilde{\alpha}} \circ H
\eeq
for any map $H: X \to \Wcal$ with arbitrary $X$, and introduce the coordinate-dependent square bracket notation
\begin{align}
&U[\bar{\gamma};c^\pi] = U, \\
& U_\Lambda[c^\pi] = U_\Lambda
\end{align}
for any map $U: \Wcal \to X$.
Using these definitions, we can for instance write \eqref{parametrisation} as
\beq
s[\omega] = S_\omega[\bar{\gamma};c^\pi].
\eeq

Continuing with physics, a change of variables in the integral in~\eqref{parametrisation} gives
\beq
S[\gamma^{\Lambda};g,\omega] = S[\gamma^{t\Lambda}; D_t g,D_t\omega].
\eeq
Therefore, given an action functional $S$, the non-trivial component of its associated chart $c$ satisfies the relation 
\beq
c^{\pi}_{t\Lambda}(s_t) = D_t c^{\pi}_{\Lambda}(s). \label{gdilatation}
\eeq
A given chart $c$ on $\Wcal$ induces a set of scale-dependent charts on the quotient space, $\mathbf{c}_\Lambda:\Mcal \to \Ccal$, defined by 
\beq
(\mathbf{c}_\Lambda(\mathbf{s}),\Lambda) = c \circ \rho_\Lambda (\mathbf{s}).
\eeq
and fulfilling the relation
\beq
\mathbf{c}_{t\Lambda}=D_t \mathbf{c}_\Lambda . \label{dilatation}
\eeq

We will also work in the coordinate basis $\{ \d^c_{\tilde{\alpha}} \}$ of T$\Wcal$ in the tangent space at each point. Given any real function $F$ in $\Wcal$, 
\beq
\d^c_{\tilde{\alpha}} F = \frac{\delta F\circ c^{-1}}{\delta g^{\tilde{\alpha}}}.\label{CoordinateVector}
\eeq
We can then write vector fields $v$ in $\mathrm{T} \,\Wcal$ as 
\begin{align}
\left. v \right|_{(s,\Lambda)} & = v^{\tilde{\alpha}} (s,\Lambda) \left. \d^c_{\tilde{\alpha}} \right|_{(s,\Lambda)} \nn
& = v^\alpha_\Lambda(s) \left. \d_\alpha^{c} \right|_{(s,\Lambda)} + v^\wedge_\Lambda(s) \left. \d^c_\wedge \right|_{(s,\Lambda)} . \label{vectorcomponents}
\end{align}
The components in this basis are given by, $v^{\tilde{\alpha}}  = v c^{\tilde{\alpha}}$.  As explained above, a vector $\left. \Ocal \right|_{(s_0,\Lambda)}= \Ocal^\alpha (s_0,\Lambda) \left. \d^c_\alpha \right|_{(s_0,\Lambda)}$ is associated to an operator (a functional of the quantum fields). In coordinates,
\begin{equation}
\left. \Ocal \right|_{(s_0,\Lambda)}[\omega]  = \Ocal^{\alpha}(s_0,\Lambda) \left. \frac{\delta  S[\gamma^\Lambda;g,\omega]}{\delta g^\alpha} \right|_{g_0}, \label{explicitO}
\end{equation}
with $g_0=c_\Lambda(s_0)$ and $S$ the action functional associated to $c$.
If the components $\Ocal^\alpha$ (with upper indices and not to be confused with the operators themselves) are of the form $\Ocal^{ax}=\sum_{n=0}^m \Ocal^{a(n)} \d^{2n}_x \delta(x-y)$, for some space-time point $y$, the operator will be local.
This is the case of the local operators associated to the basis vectors $\d^c_\alpha|_{(s,\Lambda)}$, which, as can be seen in~\refeq{explicitO} with $(\Ocal_\alpha)^{\alpha_1} = \delta^{\alpha_1}_\alpha$, depend on $\Lambda$ only through the metric, 
\begin{align}
\left. \d_\alpha^c \right|_{(s,\Lambda)} S_\omega &= \left.\Ocal_\alpha\right|_{(s,\Lambda)}[\omega] \nn
& = \Ocal^{(s)}_\alpha[\gamma^\Lambda;\omega]. \label{BasisOperator}
\end{align}
We will make extensive use of quasilocal changes of coordinates $c \to c^\prime$, given by $\zeta^{\alpha}[\bar \gamma,c^{\pi}]={c^\prime}^{\alpha}$. 
The induced changes of coordinates in the quotient space are $\boldsymbol{\zeta}_\Lambda = \mathbf{c}^\prime_\Lambda \circ \mathbf{c}_\Lambda^{-1}$. 
The vector components in~\refeq{vectorcomponents} transform in the usual way under a change of coordinates:
\beq
v^{\prime\,\tilde{\alpha}} = v^{\tilde{\alpha}_1} \d^c_{\tilde{\alpha}_1} {c^\prime}^{\tilde{\alpha}}.  
\eeq
The fact that this transformation mixes in general the $\Ccal$ and $\mathbb{R}^+$ components of the vectors, with
\beq
v^{\prime\, \alpha} = v^{\alpha_1} \d^{c}_{\alpha_1} {c^\prime}^\alpha + v^\wedge \d^c_\wedge {c^\prime}^\alpha,
\label{vectorcomponent}
\eeq
will be relevant below.


\section{Exact RG flows}
\label{sec:RG}

There exists at least a further and more interesting redundancy in the description of regularised quantum field theories.  
Given a Wilson action $s_0$ and a cutoff $\Lambda_0$, consider a new cutoff $\Lambda<\Lambda_0$ and let the new action $s$ be defined by integrating out the quantum degrees of freedom between $\Lambda$ and $\Lambda_0$:
\beq
e^{-s[\omega]} = \int \left[\Dcal \omega \right]^{\Lambda_0}_{\Lambda} \,e^{-s_0[\omega]}. \label{integrateout}
\eeq
The notation in the measure indicates that the path integral is performed with a UV cutoff $\Lambda_0$ and an IR cutoff $\Lambda$, satisfying $\left[\Dcal \omega \right]^\Lambda \left[\Dcal \omega \right]^{\Lambda_0}_{\Lambda} = \left[\Dcal \omega \right]^{\Lambda_0}$. Although we use the same symbol $\omega$ on the left-hand and right-hand sides of~\refeq{integrateout}, the action $s$ depends only on the degrees of freedom in $\omega$ that have not been integrated out. By construction, the actions $s$ and $s_0$ satisfy
\beq
Z_{\Lambda}(s)=Z_{\Lambda_0}(s_0) . \label{invariantZ} 
\eeq
We define the flow in theory space $f_t:\Wcal \to \Wcal$ such that
\beq
(s,\Lambda) = f_{\Lambda/\Lambda_0}(s_0,\Lambda_0), 
\eeq
with $f_1=\mathds{1}$. 
In general, $s \neq (s_0)_{\Lambda/\Lambda_0}$, so~\refeq{invariantZ} relates different points in $\Mcal$, as defined in the previous section: if $\mathbf{s}=[(s,\Lambda)]$ and $\mathbf{s}_0 = [(s_0,\Lambda_0)]$, 
\beq
\mathbf{Z}(\mathbf{s}) = \mathbf{Z}(\mathbf{s}_0), \label{invariantZtheory}
\eeq
The property of exact RG invariance is given by \refeq{invariantZ} and \refeq{invariantZtheory}. The latter defines the RG flow $\mathbf{f}_t$ in $\Mcal$:
\begin{align}
\mathbf{s} & = \mathbf{f}_{\Lambda/\Lambda_0}(\mathbf{s}_0) \nn
& = [f_{\Lambda/\Lambda_0}(s_0,\Lambda_0)] ,
\end{align}
with $\mathbf{f}_1=\mathds{1}$. This is a good definition, independent of the representative, since 
$f_t(s_{t^\prime},\Lambda t^\prime) = (f_t(s,\Lambda) )_{t^\prime}$, where $(s,\Lambda)_t\equiv (s_t,\Lambda t)$.
There is also an inverse relation,
\beq
f_t(s,\Lambda) = \rho_{t\Lambda} \circ \mathbf{f}_t ([(s,\Lambda)]).
\eeq
These flows are generated by beta vector fields, which are tangent to the corresponding curves. They act on any real function $F$ on  $\Wcal$ and $\Mcal$ as
\begin{align}
&\beta F =  \left. t \d_t F \circ f_{t} \right|_{1} , \\
&\boldsymbol{\beta} F = \left. t \d_t F \circ \mathbf{f}_{t} \right|_{1} 
\end{align}
respectively.
They can be used to write
the Callan-Symanzik equations
\begin{align}
&\beta Z = 0, \\  
&\boldsymbol{\beta} \mathbf{Z} = 0,
\end{align}
which are the infinitesimal versions of ~\refeq{invariantZ} and~\refeq{invariantZtheory}, respectively.
The usual description of RG flows follows once a coordinate system $c$ is chosen in $\Wcal$,
\begin{align}
f^\alpha_{t,\Lambda}[g] &=  f^\alpha_t[\gamma^\Lambda;g] \nn
& = c^\alpha \circ f_t (s,\Lambda) , \label{cf}
\end{align}
with $c^\pi(s)=g$, which agrees with our bracket notation. In local quantum field theory these flows are position-dependent quasilocal functionals of the couplings $g$, thanks to the IR cutoff in~\refeq{integrateout}. Similarly, the flows in $\Mcal$ can be parametrised as
\begin{align}
&\mathbf{f}^\alpha_{t} = \mathbf{c}_1^\alpha  \circ \mathbf{f}_t,\\
&\mathbf{f}^\alpha_{t}[\mathbf{c}_1^\pi] = \mathbf{f}^\alpha_{t}.
\end{align}
Their relation with the flows of couplings in~\refeq{cf} follows from~\refeq{dilatation}:
\beq
f^{a}_{t,\Lambda} = D_{t\Lambda} \mathbf{f}_{t}^a \left[D_{\Lambda^{-1}} c^{\pi}\right]. \label{frelation}
\eeq
The beta vector fields can be written in the coordinate basis associated to $c$:
\begin{align}
\left. \beta \right|_{(s,\Lambda)} &= \beta^{\tilde{\alpha}}(s,\Lambda) \left. \d^c_{\tilde{\alpha}}  \right|_{(s,\Lambda)} \nn
                   &=  \beta^\alpha (s,\Lambda) \left. \d^c_\alpha \right|_{(s,\Lambda)}+ \Lambda \left. \d^c_\wedge \right|_{(s,\Lambda)}.
                   \label{betacoordinates}
\end{align}
Note that the components are given by
\beq
\beta^{\tilde{\alpha}} = \left. t\d_t f^{\tilde{\alpha}}_t \right|_{1} .
\eeq
These beta functions are also quasilocal functionals of the couplings,
\beq
\beta_\Lambda^\alpha[g]= \beta^\alpha[\gamma^\Lambda;g].
\eeq 
In coordinates, the Callan-Symanzik equation has the more familiar form \footnote{Separating the vacuum energy coupling from the rest we have
\begin{align}
 \left[\Lambda \frac{\d}{\d \Lambda} + \beta_\Lambda^{\alpha^{\prime}}[g] \frac{\delta}{\delta g^{\alpha^{\prime}}} -\int d^d x\sqrt{|\gamma^{\Lambda}|}\mathcal{A}(x)\right] Z_{\Lambda}[g] = 0, \label{splitCallanSymanzik}
\end{align}
where $\alpha^{\prime}=ax$, with $a\neq 0$ and $\mathcal{A}(x)=\beta^{0x}$. Therefore, $\beta^{0x}$ represents the conformal anomaly of the theory, which is a generic feature in the presence of space-time dependent couplings.
}
\beq
\left[\Lambda \frac{\d}{\d \Lambda} + \beta_\Lambda^\alpha[g] \frac{\delta}{\delta g^\alpha} \right] Z_\Lambda[g] = 0. 
\label{CallanSymanzik}
\eeq
Finally, the beta functions in $\Mcal$ are given by the components of $\boldsymbol{\beta}$ under the chart $\mathbf{c}_1$:
\begin{equation}
\boldsymbol{\beta} = \boldsymbol{\beta}^\alpha \d^{\mathbf{c}_1}_\alpha,
\end{equation} 
and satisfy
\begin{align}
&\boldsymbol{\beta}^{\alpha} \circ \mathbf{c}_1 = \boldsymbol{\beta} \mathbf{c}_1^{\alpha},\\
&\boldsymbol{\beta}^\alpha = \left. t\d_t \mathbf{f}^\alpha_t \right|_{1}.
\end{align}
Using~\refeq{frelation}, we find the relation~\cite{Lizana:2015hqb}
\beq
\beta^a_\Lambda = D_\Lambda \boldsymbol{\beta}^a [D_\Lambda^{-1} c^{\pi}_\Lambda] + D c^{a}_\Lambda,
\eeq
with $D=\left( n^d_{(a)}-n^u_{(a)} \right)+x^\mu \d_\mu$, being $n^u_{(a)}$ ($n^d_{(a)}$) the number of contravariant (covariant) indices of $c^a$.
Under a change of coordinates $c \to c^{\prime}=\zeta[\bar\gamma; c^{\pi}]$, 
the beta functions transform into
\beq
\beta^{\prime\,\alpha} =\beta^{\alpha_1} \, \partial^c_{\alpha_1} c^{\prime\,\alpha}+
2 \bar \gamma \frac{\d}{\d \bar \gamma} \zeta^{\alpha}\left[\bar \gamma;c^{\pi}\right]. \label{betatransform}
\eeq
Notice the appearance of an inhomogeneous term, in agreement with~\refeq{vectorcomponent}.

The fixed points $\mathbf{s}_*$ of the quotient-space RG flows, with $\boldsymbol{\beta}_{\mathbf{s}}=0$, describe scale-invariant physics. In the space $\Wcal$, they correspond to points $(s_*^\Lambda,\Lambda)=\rho_\Lambda(\mathbf{s}_*)$ with trivial RG evolution $f_t(s_*^\Lambda,\Lambda) =((s^\Lambda_*)_{t},\Lambda t)$. In our parametrisations, this translates into the trivial running $g\to D_t g$. We will only consider the usual translationally invariant fixed points, with constant scalar couplings, which are thus invariant under this rescaling. In the following, we study perturbatively the theories obtained by small deformations of a given fixed point. In doing this, it can be useful to consider as well points at a finite distance, as we will see.  The set of points that flow into the fixed point under direct RG evolution is called the {\em critical} manifold, while the set of points reaching the fixed point under inverse evolution describes ``renormalised trajectories''. The latter can be used to define scale-dependent renormalised theories in the continuum limit~\cite{Morris:1998da}. 


\section{Normal coordinates}
\label{sec:normal}

In this section we single out a special set of coordinates, valid in some region around a given fixed point, in which the beta functions and RG flows take a remarkably simple form. Later on we will see that these coordinates are closely related to the process of renormalisation.

We start with an arbitrary chart $c$ with the fixed point of interest located at $c^\pi(s_*^\Lambda,\Lambda)=g_*=0$. That is, $\beta^\alpha_\Lambda[g_*]=0$.  Close to this fixed point, the beta functions can be expanded as
\begin{equation}
\beta^{\alpha} = \beta_{\alpha_1}^{\alpha}(\bar \gamma) c^{\alpha_1}+ \sum_{n\geq2}  \beta^{\alpha}_{\alpha_1...\alpha_n} (\bar \gamma) c^{\alpha_1}...\,c^{\alpha_n}.  \label{betaexpansion}
\end{equation}
Both sides of this equation are maps on $\Wcal$.
In a local quantum field theory the beta functions are local or quasilocal, in the sense that they can be written as a finite or infinite sum of products of Dirac delta functions and their derivatives, contracted if necessary with the (inverse) metric.
Consider first the linearised approximation and let $\lambda_{(a)}$ be eigenvalues of $\beta^{\alpha_1}_{\alpha_2}$, which we assume to be real numbers.
The linear part of the beta function can be maximally aligned with the couplings by a linear reparametrisation 
\begin{align}
\bar c^\alpha  
 = \zeta^{\alpha}_{\alpha_1} (\bar \gamma) c^{\alpha_1}, \label{linearUV} 
\end{align}
with quasilocal $ \zeta^{\alpha}_{\alpha_1} (\gamma)$, which puts the linear part of the beta function in a generalised Jordan form,
\beq
\bar{\beta}^\alpha = - \lambda^{\alpha}_{\alpha_1}(\bar \gamma) \bar{c}^{\alpha_1} + O(\bar{c}^2),\label{BetaFunctionLin}
\eeq
where the quasilocal matrix $\lambda^{\alpha}_{\alpha_1}$ has, neglecting metrics, a diagonal block structure, with each block having a unique eigenvalue $\lambda_{(a)}$. 
Non-vanishing terms with $n_\d$ derivatives in off-block positions $(a,a_1)$ are only allowed when
\begin{equation}
\left[\lambda_{(a)}-n^u_{(a)}+n^d_{(a)}\right] - \left[\lambda_{(a_1)} -n^u_{(a_1)}+n^d_{(a_1)}\right] = n_\d, \label{GenJordanCond}
\end{equation} 
where we have allowed the directions $a$, $a_1$ to have tensor character, with $n^u_{(a_i)}$ ($n^d_{(a_i)}$) the number of contravariant (covariant) indices, in coordinates.
The number of derivatives and tensor indices enters this condition through the non-homogenous term in~\refeq{betatransform}.
Notice that, by covariance, there is a relation between $n^u_{(a_i)}$, $n^d_{(a_i)}$, $n_{\d}$ and the number of metrics $n_{\left(\gamma\right)}$ and inverse metrics $n_{\left(\gamma^{-1}\right)}$ appearing in $\lambda^{\alpha}_{\alpha_1}$:
\begin{equation}
2n_{\left(\gamma^{-1}\right)}-2n_{\left(\gamma\right)}=n^u_{(a)}-n^d_{(a)}-n^u_{(a_1)}+n^d_{(a_1)}+n_{\d}.\label{ConstraintNs}
\end{equation} 
The eigenvalues $\lambda_{(a)}$ give the (quantum) scaling dimensions $\Delta_{(a)}$ of the eigendeformations of the fixed-point theory: $\Delta_{(a)}=d-\lambda_{(a)}+n^u_{(a)}-n^d_{(a)}$. By definition these dimensions are less than, equal to and greater than $d$ for relevant, marginal and irrelevant deformations, respectively. Usually, the number of relevant eigenvalues is finite~\cite{Morris:1998da}. In a unitary CFT, the eigendeformations span the complete space $\mathcal{I}$ and $\lambda$ can be written in completely diagonal form. Nevertheless, thinking of the possible application to logarithmic CFT, we shall proceed in the general case without the assumption of diagonalizability. 

In going beyond the linear approximation, it is important to distinguish certain exceptional cases. The set of eigenvalues $\{\lambda_{(a)}\}$ is said to be {\em resonant} if
\beq
\sum_{i=1}^m \left[\lambda_{(a_i)}- n^u_{(a_i)} + n^d_{(a_i)} \right]+ n_\d = \lambda_{(a)} - n^u_{(a)} + n^d_{(a)} \label{resonance}
\eeq
for some (possibly repeated) eigendirections $a$, $a_1,\ldots,a_m$, with $n_\d$ a non-negative integer and $m\geq 2$. Relations between eigenvalues of the form \refeq{resonance} are called resonances and the eigenvalue $\lambda_{(a)}$ is said to be composite. Scalar marginal directions, i.e. $\lambda_{(a_i)}=0$ for some $a_i$, imply an infinite number of resonances and that all eigenvalues are composite. Note that the condition for non-negative off-diagonal terms in the linear part has the same form as \refeq{resonance}, with $m=1$. We will say that the eigenvalues, or the associated dimensions, are {\em exceptional} when the relations \refeq{resonance} occur for some $m\geq 1$. Non-exceptional eigenvalues or dimensions will be called {\em generic}. For non-resonant eigenvalues, by the Poincar\'e linearisation theorem~(see, e.g.~\cite{linearization}) we know that, at least as a formal series, we can find a coordinate transformation such that in the new coordinates the 
beta function is linear:
\beq
\bar{\beta}^\alpha = - \lambda^{\alpha}_{\alpha_1}(\bar \gamma) \bar{c}^{\alpha_1} ~~\mbox{(non-resonant)}. \label{nonresonantbeta}
\eeq
In this case, the integration of this vector field is trivial and the RG flows are given by 
\beq
\bar{f}^\alpha_{t} = \mathcal{P}\exp \left\{-\int_{1}^{t} \frac{d t'}{t'} \lambda \left(t'^2\bar \gamma\right) \right\} ^{\alpha}_{\alpha_1} \bar{c}^{\alpha_1}, \label{linearJordanflow}
\eeq
where $\mathcal{P} \exp $ is the path-ordered exponential. The flows can also be written in a more useful manner:
\begin{equation}
\bar{f}^\alpha_{t} =  t^{-\lambda_{(\alpha)}} \left[\mathcal{M}_{t}(\bar \gamma)\right]^{\alpha}_{\alpha_1} \bar{c}^{\alpha_1},\label{genericflow}
\end{equation} 
where $\mathcal{M}_{t}(\gamma)$ is the identity matrix in a fully diagonalisable case and depends logarithmically in $t$ otherwise. In all cases, $\mathcal{M}_1=\mathds{1}$. It satisfies the same requirements as $\lambda(\gamma)$: it is diagonalized in subspaces with the same eigenvalue $\lambda_{(\alpha)}$ and can have non-vanishing terms with $n_\d$ derivatives in off-block positions $(a,a_1)$ only when~\refeq{GenJordanCond} is satisfied.
To prove~\refeq{genericflow}, let us introduce it in the linear differential flow equation:
\begin{align}
t \frac{\partial}{\partial t}\bar{f}^\alpha_{t}=& 
t ^{-\lambda_{(\alpha)}} \left\{ - \lambda_{(\alpha)}\left[\mathcal{M}_{t}(\bar \gamma)\right]^{\alpha}_{\alpha_2}
+t\frac{\partial}{\partial t}\left[ \mathcal{M}_{t} (\bar \gamma)  \right]^{\alpha}_{\alpha_2} \right\}\bar{c}^{\alpha_2}\nn
=&-\lambda^{\alpha}_{\alpha_1}\left(t^2\bar \gamma\right) t^{-\lambda_{(\alpha_1)}} \left[\mathcal{M}_{t}(\bar \gamma)\right]^{\alpha_1}_{\alpha_2} \bar{c}^{\alpha_2}\nn
=&-t^{-\lambda_{(\alpha)}} \lambda^{\alpha}_{\alpha_1}(\bar \gamma)  \left[\mathcal{M}_{t}(\bar \gamma)\right]^{\alpha_1}_{\alpha_2} \bar{c}^{\alpha_2}.
\end{align}
In the second line we have used the linear form of the beta function \refeq{BetaFunctionLin} and in the third one we have commuted the first two matrices, which is allowed by the specific form of the matrix $\lambda(\gamma)$. Then, combining the first and third equations, we obtain
\begin{equation}
t\frac{\partial}{\partial t}\left[ \mathcal{M}_{t} (\gamma)  \right]^{\alpha}_{\alpha_1} =-\left[ \lambda^{\alpha}_{\alpha_2}(\gamma)-\lambda_{(\alpha)}\delta^{\alpha}_{\alpha_2} \right]
\left[ \mathcal{M}_{t} (\gamma)  \right]^{\alpha_2}_{\alpha_1}.
\end{equation} 
This is a linear and autonomous differential equation.  The solution is found to be
\begin{equation}
\left[ \mathcal{M}_{t} (\gamma)  \right]^{\alpha}_{\alpha_1} =\exp \left\{-\left[ \lambda^{\alpha}_{\alpha_1}(\gamma)-\lambda_{(\alpha)}\delta^{\alpha}_{\alpha_1} \right] \log t \right\}.
\end{equation} 
The matrix $\left[ \lambda^{\alpha}_{\alpha_2}(\gamma)-\lambda_{(\alpha)}\delta^{\alpha}_{\alpha_2} \right]$ has only vanishing eigenvalues, i.e.\ it is idempotent. Therefore, the Taylor expansion of the exponential above has only a finite number of logarithmic terms in $t$.

\begin{figure}[t!]
\begin{center}
\includegraphics[width=4cm]{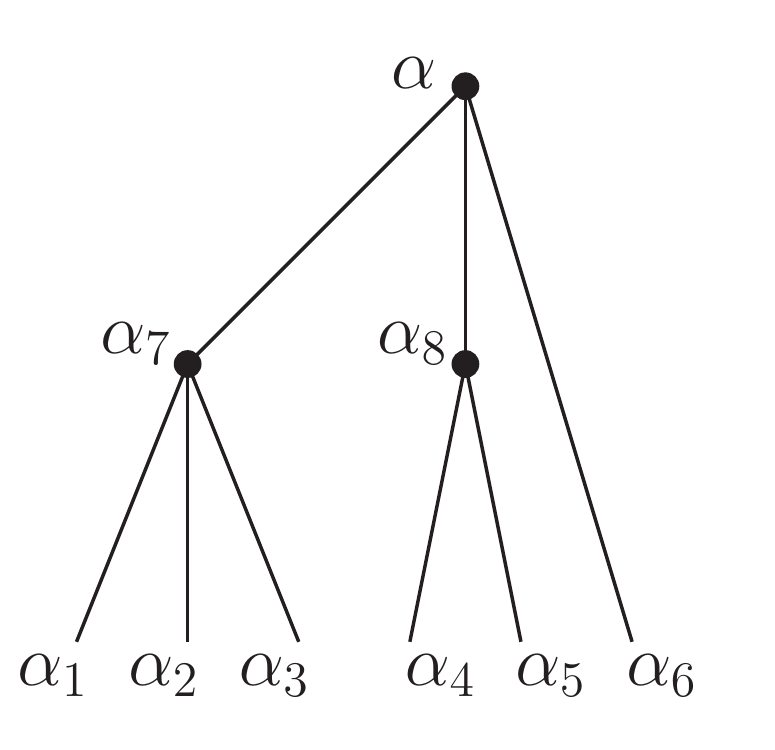}
\caption{Diagrammatic representation of the contribution to $[B_p]^{\alpha}_{\alpha_1\alpha_2\alpha_3\alpha_4\alpha_5\alpha_6}$ written in~\refeq{tree}: $\beta^{\alpha_7}_{\alpha_1 \alpha_2\alpha_3}\beta^{\alpha_8}_{\alpha_4 \alpha_5}  \beta^{\alpha}_{\alpha_7 \alpha_8 \alpha_6} $. Each dot represents a coefficient $\beta$ of the product. The index next to a dot is the upper index of the associated coefficient $\beta$, while the indices to which the dot is linked to, downwards,  are the lower indices of the associated coefficient $\beta$.} 
\label{fig:tree}
\end{center}
\end{figure}

When the set of eigenvalues is resonant, complete linearisation is not possible in general. However, the Poincar\'e-Dulac theorem~(see e.g.~\cite{PoincareDulac}) implies, at least in the sense of formal power series, that we can choose coordinates in which the beta functions take the {\em normal form}\footnote{Actually, we are generalising the Poincar\'e-Dulac theorem to the case of quasilocal vector fields in a space of functions. To prove this generalisation, at least at a finite order in the coupling and derivative expansions, we can simply choose a space-time point and treat the $n$-derivatives of couplings at that point as independent couplings.}
\beq
\bar{\beta}^{ \alpha} = -\lambda^\alpha_{\alpha_1}(\bar \gamma)  \bar{c}^{ \alpha_1}+ \sum_{n\geq2} \bar{\beta}^{ \alpha}_{ \alpha_1...\alpha_n}(\bar \gamma)\bar{c}^{\alpha_1}...\,\bar{c}^{\alpha_n}, \label{resonantbeta}
\eeq
where $\bar{\beta}^{ax}_{a_1x_1\ldots a_m x_m}$ has support at $x_1=\ldots=x_m$ and is non-vanishing only when condition~\refeq{resonance} is met for $n_\d$ equal to the number of derivatives in it. Again, further simplifications are admitted, but the form~\refeq{resonantbeta} will be sufficient for our purposes. Obviously, \refeq{resonantbeta} reduces to \refeq{nonresonantbeta} for non-resonant eigenvalues.
Analogously to~\refeq{ConstraintNs}, by covariance, the number of metrics $n_{(\gamma)}$ and inverse metrics $n_{\left(\gamma^{-1}\right)}$ of $\bar{\beta}^{ \alpha}_{ \alpha_1...\alpha_n}(\gamma)$ is constrained by
\begin{equation}
2n_{\left(\gamma^{-1}\right)}-2n_{\left(\gamma\right)}=n^u_{(a)}-n^d_{(a)}-\sum_{i=1}^n \left(n^u_{(a_i)}-n^d_{(a_i)}\right)+n_{\d}.\label{ConstraintNs2}
\end{equation} 
The coordinates in which the beta functions take the form~\refeq{resonantbeta} will be called {\em normal coordinates}. They are not unique. 
In normal coordinates, the RG flows have the perturbative form
\begin{align}
\bar{f}^{\alpha}_{t} =&t^{-\lambda_{(\alpha)}} \left\{\bar{c}^{\alpha} +\sum_{m=1}^\infty \left[\sum_{p=1}^{p^{\mathrm{max}}_{\alpha_1\cdots\alpha_m}} \log^p  t   \left[B_p\right]^{\alpha}_{ \alpha_1... \alpha_m}(\bar \gamma)\right]\bar{c}^{ \alpha_1}\cdots \bar{c}^{ \alpha_m}\right\} \nn
=& t^{-\lambda_{(\alpha)}} \bar{c}^{\alpha} +\sum_{m=1}^\infty \left[\sum_{p=1}^{p^{\mathrm{max}}_{\alpha_1\cdots\alpha_m}} \log^p  t   \left[B_p\right]^{\alpha}_{ \alpha_1... \alpha_m}\left(t^2\bar \gamma\right)\right]\,\left[t^{-\lambda_{(\alpha_1)}}\bar{c}^{ \alpha_1}\right]\cdots \left[t^{-\lambda_{(\alpha_m)}}\bar{c}^{ \alpha_m}\right],  \label{resonantflow}
\end{align}
where $[B_1]^\alpha_{\alpha_1\ldots \alpha_m}=\bar{\beta}^\alpha_{\alpha_1\ldots \alpha_m}$, with $\bar \beta^\alpha_{\alpha_1}=\lambda_{(\alpha_1)}\delta^{\alpha}_{\alpha_1}-\lambda^{\alpha}_{\alpha_1}$. The functions $[B_p]^{ax}_{a_1x_1\ldots a_m x_m}$ with $p>1$ have also support at $x_1=\ldots=x_m=x$. They can be computed (up to combinatorial coefficients) by summing all the products of $p$ functions $\bar\beta^{\alpha}_{\alpha_1...\alpha_r}$, $r\geq1$, with upper indices contracted with lower indices in such a way that the only free upper and lower indices are $ax$ and $a_1x_1\ldots a_m x_m$ respectively. For instance,
\begin{equation}
[B_3]^{\alpha}_{\alpha_1\alpha_2\alpha_3\alpha_4\alpha_5\alpha_6}(\gamma)= \dots + \frac{1}{S_T} 
\beta^{\alpha_7}_{\alpha_1 \alpha_2\alpha_3}(\gamma)\beta^{\alpha_8}_{\alpha_4 \alpha_5} (\gamma)  \beta^{\alpha}_{\alpha_7 \alpha_8 \alpha_6} (\gamma)  + \dots
\label{tree}
\end{equation}
where $S_T$ is a combinatorial coefficient.
Each contribution of this type can be seen as a tree $T$, whose elements are the coefficients $\beta^{\alpha}_{\alpha^{\prime} \dots \alpha^{\prime \prime}}(\gamma)$ of the product, and the tree structure is given by the contraction of the indices. For example, the contribution written explicitly in~\refeq{tree} is the tree $T$ given by the set
$\left\{\beta^{\alpha_7}_{\alpha_1 \alpha_2\alpha_3},\, \beta^{\alpha_8}_{\alpha_4 \alpha_5},\,  \beta^{\alpha}_{\alpha_7 \alpha_8 \alpha_6} \right\}$ and represented in Fig.~\ref{fig:tree}.  Each $[B_p]^\alpha_{\alpha_1\dots \alpha_n}$ is a sum of the contributions of all the possible trees with $p$ dots that connect $\alpha$ (on top) with $\alpha_1 \dots \alpha_n$ (at bottom).

Terms in $[B_p]^\alpha_{\alpha_1\ldots \alpha_m}$ with $n_\d$ derivatives are non-vanishing only when the resonant condition~\refeq{resonance} is satisfied for the set of the eigenvalues $\{\alpha_1\ldots \alpha_m\}$ and $\alpha$. The constraint of~\refeq{ConstraintNs2} also holds.
Moreover, for a given order in the number of couplings, $\alpha_1\ldots \alpha_m$, the sum in $p$ is finite and stops at some finite $p^{\mathrm{max}}_{\alpha_1\cdots\alpha_m}$, depending on the order. This is because $\bar \beta^{\alpha}_{\alpha_1}$ is nilpotent and thus the number of possible trees to construct $[B_p]^\alpha_{\alpha_1\ldots \alpha_m}$ is finite.
Logarithmic differentiation of~\refeq{resonantflow} with respect to $t$ gives the beta function~\refeq{resonantbeta} order by order in $\bar{c}$. For generic dimensions, \refeq{resonantflow} reduces to~\refeq{genericflow}. Note that, non-trivially, $\bar{f}_{t}$ is the inverse of $\bar{f}_{t^{-1}}$, as it should. Observe also that, up to possible log terms and factors of the metric, the components  $\bar{f}_{t}^\alpha$ scale homogeneously as a power of $t$. This simple feature is no longer apparent when we write the flows in other coordinates.


\section{Correlation functions}
\label{sec:Correlation}

\subsection{Renormalisation}

A central quantity of interest in quantum field theory is the continuum limit of correlation functions, evaluated at cutoff-independent space-time points. 
Given a chart $c$ and a point $(s,\Lambda)\in \Wcal$, we define the bare (or cutoff) $n$-point connected correlation functions as
\begin{align}
G^{(s,\Lambda)}_{\alpha_1\ldots\alpha_n}& = \left. \d^c_{\alpha_1}\ldots \d^c_{\alpha_n} \right|_{(s,\Lambda)} W \nn
& =
\left.\frac{\delta ^{n} W_{\Lambda}[ g]}{\delta g^{\alpha_1} \ldots \delta g^{\alpha_n}}\right|_{c(s,\Lambda)},
\label{barefunction1}
\end{align} 
where the generator $W$ is given
\beq
Z = e^{W}.
\eeq
The symbol $\d^c_{\alpha_1}\ldots \d^c_{\alpha_n}|_{(s,\Lambda)}$ on the right-hand side of the first line of \refeq{barefunction1} refers to the sequential action of the vector fields $\d_{\alpha_i}$ associated to the coordinates $c$ on the function $W$, eventually evaluated at $(s,\Lambda)$. These vector fields commute among themselves, so the correlators do not depend on the order of the operators. The corresponding basis vectors at $(s,\Lambda)$ are identified with the local operators $\Ocal_{\alpha_i}(s,\Lambda)$ as given in~\refeq{BasisOperator}. The definition~\refeq{barefunction1} is a convenient generalisation of the standard definition with linear sources. In this paper we will be ultimately interested in correlation functions at a fixed point $\mathbf{s}_*$. 
Obviously, this definition of the correlation function is coordinate dependent, even if are not indicating the chart explicitly. It is important to note that for $n\geq 2$ the
correlators~\refeq{barefunction1} transform  nonlinearly under changes of coordinates in $\Wcal$. For instance, the {\em same} 2-point correlation function written in the basis associated with a different coordinate system $c^\prime$ would read
\begin{align}
G^{(s,\Lambda)}_{\alpha_1\alpha_2}  = & \left. 
(\d^c_{\alpha_1} {c^\prime}^{\alpha_3}) (\d^c_{\alpha_2} {c^\prime}^{\alpha_4})
\d^{c^\prime}_{\alpha_3} \d^{c^\prime}_{\alpha_4} \right|_{(s,\Lambda)} W \nn
& \mbox{} + \left. (\d^c_{\alpha_1} {c^\prime}^{\alpha_3}) (\d^c_{\alpha_3}\d^c_{\alpha_2} {c^\prime}^{\alpha_4}) \d^{c^\prime}_{\alpha_4} \right|_{(s,\Lambda)} W . \label{examplecoc}
\end{align}
The continuum limit $\Lambda \to \infty$ of these correlators is divergent in non-trivial theories. Therefore, renormalisation is necessary. Let us introduce a family of {\em renormalisation charts} 
$r_t: \Wcal \to \Ccal\times \mathbb{R}^+$, with $t\in \mathbb{R}^+$, of the form
$r_t(s,\Lambda) = (r^{\pi}_{t,\Lambda}(s),\Lambda/t)$. We fix the origin of these special coordinates such that, for all $t$,
\beq
r^{\pi}_{t}(s_c^{\Lambda},\Lambda)= 0 \label{criticalpoint}
\eeq
where the ``critical point'' $\mathbf{s}_c=[(s^\Lambda_c,\Lambda)]$ is an arbitrary point on the critical manifold attracted by the fixed point $\mathbf{s}_*$.\footnote{If we were to calculate correlators at a different point in the renormalised manifold we would instead need ${r_t^{-1}}(0,\mu)$ to approach $(s_c^{t \mu},t \mu)$ at a tuned rate.} 
Let $h_t = r^{-1}_t$  be the inverse of the renormalisation chart and let us write $h_t(g_R,\mu) = (h_{t \mu}(g_R),t\mu)$. The $\mu$-dependent maps $h^\alpha_{\Lambda}(g_R) = c^\alpha \circ h_{\Lambda}(g_R)$ play the role of {\em bare couplings} (see~\cite{Lizana:2015hqb}). In a local quantum field theory, the bare couplings can be chosen to be local functionals of the {\em renormalised couplings} $g_R^a$.
We write the basis vectors associated to the renormalisation charts as $\d_\alpha^{r_t}$. 
They can be related to local {\em renormalised operators} $[\Ocal^t_\alpha]$ by
\begin{align}
\left. \d_\alpha^{r_t} \right|_{(s,\Lambda)} S_\omega & =  \left. [\Ocal^t_\alpha] \right|_{(s,\Lambda)}[\omega]  \nn 
& =[\Ocal^t_\alpha]^{(s)}[\gamma^{\Lambda};\omega] .
\end{align} 
We are now ready to define the renormalised correlation functions as the continuum limit of the correlators of renormalised operators:
\begin{align}
G^R_{\alpha_1\ldots\alpha_n}&=\lim_{t \to\infty}  \left. \d^{r_t}_{\alpha_1}\ldots \d^{r_t}_{\alpha_n} \right|_{(s^{t \mu}_c,t \mu)} W \nn
& =
\lim_{\Lambda \to\infty}
\left. \frac{\delta ^{n} W_{\Lambda}[h_{\Lambda}(g_R)]}{\delta g_R^{\alpha_1} \ldots \delta g_R^{\alpha_n}} \right|_{0} ,
\label{renormalization}
\end{align} 
with $r_t$ (or equivalently, $h_\Lambda$) chosen in such a way that the limit is well defined (and non-trivial). There is a large degree of freedom in the choice of renormalisation scheme, i.e.\ in the choice of the particular renormalisation charts that do the job. The renormalised correlators are scheme dependent, even if many details fade in the continuum limit. In particular, they depend on the renormalisation scale $\mu$ in a way that is determined by the RG~\cite{Lizana:2015hqb}.

In actual calculations,  \refeq{renormalization} needs to be written in some specified coordinate system $c$. This involves in particular writing the vector fields $\d_\alpha^{r_t}$ in the coordinate basis $\{\d^c_\alpha\}$. As illustrated in~\refeq{examplecoc}, non-linear terms will appear under the change of coordinates $r_t \to c$. 
To keep track of those terms and to preserve covariance, let us introduce a covariant derivative $\nabla^t$ acting on tensor fields, which is characterised by being trivial (with vanishing Christoffel symbols) in the $r_t$ coordinates:
\beq
\nabla^t_{\d^{r_t}_{\alpha_1}} \d^{r_t}_{\alpha_2} = 0. \label{trivial}
\eeq
Here, $\nabla^t_v$ is the covariant derivative along the vector $v$.
In other words, by definition the covariant derivative is just an ordinary derivative in the renormalisation coordinates. 
Consequently, the corresponding connection is symmetric and flat. In arbitrary coordinates $c$, the Christoffel symbols are given by
\beq
\Gamma^{t\,\tilde{\alpha}_1}_{\tilde{\alpha}_2 \tilde{\alpha}_3} = (\d^{r_t}_{\tilde{\alpha}_4} {c}^{\tilde{\alpha}_1}) (\d^c_{\tilde{\alpha}_2} \d^c_{\tilde{\alpha}_3} r_t^{\tilde{\alpha}_4}). \label{Christoffel}
\eeq
The symbol $\Gamma^{t\, \alpha_1}_{\alpha_2 \alpha_3}$ is quasilocal in its three space-time indices.
In this language, the renormalised correlators read
\begin{align}
G^R_{\alpha_1\ldots\alpha_n}& = \lim_{t \to \infty}\, \left. \nabla^t_{\d^{r_t}_{\alpha_1}}\ldots \nabla^t_{\d^{r_t}_{\alpha_n}} W  \right|_{(s_c^{t\mu},t \mu)} \nn
& =  \lim_{t \to \infty}  \; \left. [\mathcal{O}^t_{\alpha_1}]^{\alpha_{n+1}} \ldots [\mathcal{O}^t_{\alpha_n}]^{\alpha_{2n}} 
\nabla^t_{\d^c_{\alpha_{n+1}}} \ldots \nabla^t_{\d^c_{\alpha_{2n}}} W \right|_{(s_c^{t\mu},t\mu)}  . \label{rencorr}
\end{align}
The precise meaning of these two equations deserves a short explanation. On the right-hand side of the first line, $\nabla_{\d^{r_t}_{\alpha_1}}^t \ldots \nabla_{\d^{r_t}_{\alpha_n}}^t W$ can be understood as the components of the $(0,n)$ tensor field $\nabla^t \ldots \nabla^t W$ in the $r_t$ coordinate basis. In the second line, the covariant derivatives are taken along the coordinate basis vectors associated to an arbitrary coordinate system $c$ and $\nabla^t_{\d^c_{\alpha_1}}\ldots \nabla^t_{\d^c_{\alpha_n}} W$ are to be understood as the components of the same tensor field $\nabla^t \ldots \nabla^t W$ in this later basis. The coefficients in front arise from the tensor transformation law. They are given by
\beq
[\Ocal^t_{\alpha_1}]^{\alpha_2} = \d^{r_t}_{\alpha_1} c^{\alpha_2}
\eeq
and are just the components of the renormalised operators in the $c$ basis. For local bare couplings, $[\Ocal_{a_1 x_1}]^{a_2 x_2}$ is a sum of terms proportional to $\delta(x_1-x_2)$ and derivatives of it. At this point, we can extend in a natural way the definition of renormalised correlation functions to arbitrary operators:
\beq
G^R_{\mathcal{O}_{1} \ldots \mathcal{O}_{n}} =  \lim_{t \to \infty}  \; \left. [\mathcal{O}^t_{1}]^{\alpha_{n+1}} \ldots [\mathcal{O}^t_{n}]^{\alpha_{2n}} 
\nabla^t_{\d_{\alpha_{n+1}}} \ldots \nabla^t_{\d_{\alpha_{2n}}} W \right|_{(s_c^{t\mu},t\mu)}  .
\eeq
However, in this paper we only consider the particular renormalised correlators in~\refeq{rencorr}.

The Christoffel symbols provide the non-linear counterterms that are necessary in generic coordinates. For example, the renormalised two-point functions read
\begin{align}
G^R_{\alpha_1 \alpha_2}
& =  \lim_{t \to \infty}  \;\left. [\mathcal{O}^t_{\alpha_1}]^{\alpha_{3}} [\mathcal{O}^t_{\alpha_2}]^{\alpha_{4}}  
\nabla^t_{\d^c_{\alpha_{3}}} \nabla^t_{\d^c_{\alpha_{4}}}  W \right|_{(s_c^{t\mu},t\mu)} \nn
& = \lim_{t \to \infty}  \; \left\{\left. [\mathcal{O}^t_{\alpha_1}]^{\alpha_{3}} [\mathcal{O}^t_{\alpha_2}]^{\alpha_{4}} 
 \d^c_{\alpha_{3}} \d^c_{\alpha_{4}}  W \right|_{(s_c^{t\mu},t\mu)}
- \left.  [\mathcal{O}^t_{\alpha_1}]^{\alpha_{3}} [\mathcal{O}^t_{\alpha_2}]^{\alpha_4} \Gamma^{t\, \alpha_5}_{\alpha_3\alpha_4} \d^c_{\alpha_5} W \right|_{(s_c^{t\mu},t\mu)}
\right\}.\label{TwoPointFunct}
\end{align}
We have taken into account the fact that $\Gamma^{t\; \, \wedge}_{\alpha_1 \alpha_2}=0$. The first term in the last line takes care of the non-local divergences, while
the second term cancels the local divergences that appear when the space-time points in $[\Ocal_{\alpha_1}^t]^{\alpha_3}$ and $[\Ocal_{\alpha_2}^t]^{\alpha_4}$ coincide.  From~\refeq{Christoffel},
\beq
[\mathcal{O}^t_{\alpha_1}]^{\alpha_{3}} [\mathcal{O}^t_{\alpha_2}]^{\alpha_4} \Gamma^{t\,\alpha_5}_{\alpha_3\alpha_4} = -C^{t\, \alpha_3}_{\alpha_1 \alpha_2} [\Ocal^t_{\alpha_3}]^{\alpha_5} ,\label{NonLinCounterterms}
\eeq
with the counterterms
\begin{equation}
C^{t\,\alpha}_{\alpha_1\alpha_2}= \left(\partial^{r_t}_{\alpha_1}\,\partial^{r_t}_{\alpha_2}c^{\alpha_3}\right)\left(\partial^c_{\alpha_3} r^{\alpha}_t\right)\label{QuadCounterT}
\end{equation}
quasilocal in the space-time part of their indices. Because the derivatives come with inverse metrics that decrease the degree of divergence, we can truncate the derivative expansion of the expression above and get local counterterms.  \refeq{NonLinCounterterms} used in~\refeq{TwoPointFunct} has the form of an OPE: in fact, in order to give finite continuum correlators, the singular parts in the counterterms $C^{t\,a x}_{a_1 x_1 a_2 x_2}$ must cancel the singularities for coincident points $x_1\sim x_2$ of the $[\mathcal{O}^t_{a x}]$ term in the OPE of $[\mathcal{O}^t_{a_1 x_1}]$ and $[\mathcal{O}^t_{a_2 x_2}]$.

The higher-point renormalised functions involve derivatives of the Christoffel symbols. They can also be written in terms of counterterms 
\beq
C^{t\,\alpha}_{\alpha_1\alpha_2\ldots \alpha_m}= \left(\partial^{r_t}_{\alpha_1}\,\partial^{r_t}_{\alpha_2} \ldots \partial^{r_t}_{\alpha_m}c^{\alpha^\prime}\right)\left(\partial^c_{\alpha^\prime} r^{\alpha}_t\right) \label{generalct}
\eeq
as
\begin{align}
G^R_{\alpha_1 \alpha_2 \dots \alpha_n}
& = \lim_{t \to \infty}  \sum_{p\in \Pi_n} \left. \left(\prod_{r=1}^{|p|} 
C^{t\;\alpha^\prime_r}_{\alpha_{p_r}} \right) 
\left[\mathcal{O}^t_{\alpha^\prime_1}\right]^{\alpha^{\prime\prime}_{1}} \cdots \left[\mathcal{O}^t_{\alpha^\prime_{|p|}}\right]^{\alpha^{\prime\prime}_{|p|}} 
 \d^c_{\alpha^{\prime\prime}_{1}} \ldots \d^c_{\alpha^{\prime\prime}_{|p|}}  W \right|_{(s_c^{t\mu},t\mu)}, \label{npointR}
\end{align} 
where $\Pi_n$ is the set of partitions of $\{1,2,\ldots,n\}$, $p_r$ is the $r$-th element of the partition $p$, $\alpha_{p_r}$ is a collective index given by
\beq
\alpha_{p_r} = \alpha_{p_{r1}} \ldots \alpha_{p_{r|p_r|}},
\eeq
with $p_{ri}$ the $i$-th element of the set $p_r$, and for any set $A$, $|A|$ is its cardinality. Furthermore, we have defined $C^{t\; \alpha_1}_{\alpha_2} = \delta^{\alpha_1}_{\alpha_2}$.
The counterterms in~\refeq{npointR} cancel not only local but also semilocal divergences when only a subset of points coincide~\cite{Lizana:2015hqb,PerezVictoria:2001pa,Bzowski:2015pba}. 

\subsection{Connection with renormalisation group}

Let us next connect the renormalisation process with the exact RG flows.
Using RG invariance, as given by~\refeq{invariantZ}, we rewrite~\refeq{renormalization} as
\begin{align}
G^R_{\alpha_1 \ldots \alpha_n} & =\lim_{t\to\infty}
\left. \nabla^t_{\d^{r_t}_{\alpha_1}}\ldots \nabla^t_{\d^{r_t}_{\alpha_n}}  W \circ f_{1/t}  \right|_{(s^{t \mu}_c,t \mu)}  \nn
&=
\lim_{\Lambda\to\infty}
\left.\frac{\delta ^{n} W( f_{\mu/\Lambda,\Lambda}\circ  h_{\Lambda}(g_R))}{\delta g_R^{\alpha_1}\ldots\delta g_R^{\alpha_n}} \right|_{0} , \label{correlationRG}
\end{align}
for any scale $\mu$. Since  $W(s,\mu)$ is finite for finite $s$, it is clear that the limit in \refeq{correlationRG} will be well defined as long as  $f_{\mu/\Lambda,\Lambda}\circ  h_{\Lambda}(g_R)$ stays finite when $\Lambda$ approaches infinity, at least for $g_R$ in some neighbourhood of 0. In this manner, we have rephrased the problem of removing divergences as the Wilsonian problem of finding curves $h_\Lambda(g)$ that, when composed with the RG evolution, have a well defined limit. The condition~\refeq{criticalpoint} ensures that $\lim_{\Lambda\to \infty}  f_{\mu/\Lambda,\Lambda}\circ  h_{\Lambda}(0)=s^\mu_*$. To get non-trivial correlators we need that the combined limit, besides being finite, reaches points different from $s^\mu_*$ when $g_R\neq 0$.
More precisely, the limit will be finite and non-singular if 
\beq
\lim_{t\to \infty} r_t \circ f_t = c  \label{rflimit}
\eeq
is a well-defined chart in the neighbourhood of the fixed point. 

Let us recast these statements in infinitesimal form. Because the correlation functions are given by (covariant) derivatives, for a finite set of correlators it is sufficient that the corresponding derivatives be well defined in the continuum limit. For a fixed and finite $t>0$, the RG flow $f_{t}$ defines a diffeomorphism which takes points in a region $\mathcal{A}\subset \Wcal$ onto points in a region $\mathcal{A}^t \subset \Wcal$.
This map can be used to transport any differential structure between $\mathcal{A}$ and $\mathcal{A}^t$. Recall that a vector field $v$ is transported with the differential $f^*_t:T\mathcal{A}\to T\mathcal{A}^t$,
\beq
(f^*_t v )F = v \, F\circ f_t
\eeq
for any function $F$ in $\mathcal{A}^t$, while the pullback of a one-form field $\phi$ in $T^* \mathcal{A}^t$ is given by
\beq
((f_t)_* \phi)(v) = \phi(f_t^* v) 
\eeq
for any vector field $v$ in $T\mathcal{A}$.
A tensor field $T$ of an arbitrary type $(n,m)$ can be transported from the space of tensor fields in $\mathcal{A}$ to the one in $\mathcal{A}^t$ using the pullback of $f_{t}$, $(f_t)_*$ and the differential of its inverse, $f_{1/t}^*$: 
\beq
(f^*_t T)(\phi_1,\ldots,\phi_n;v_1,\ldots v_n) = T((f_t)_* \phi_1,\ldots, (f_t)_* \phi_n;f_{1/t}^*v_1,\ldots,f_{1/t}^* v_n),
\eeq
where $\phi_i$ and $v_i$ are, respectively, dual vector and vector fields in $\mathcal{A}^t$.
Similarly, any connection $\nabla$ in $\mathcal{A}$ can be transported into another connection $\nabla^\prime=f^*_{t} \nabla$ in $\mathcal{A}^t$:
\begin{equation}
\nabla^\prime_{v} T =  f_{t}^* \left[\nabla_{ f_{1/t}^* v}   (f_{1/t}^* T)      \right],
\end{equation} 
for $v$ and $T$ arbitrary vector and tensor fields in $\mathcal{A}^t$.
Using all this, we can write \refeq{correlationRG} as
\beq
G^R_{\alpha_1 \ldots \alpha_n}  =\lim_{t\to\infty}
\left. f_{1/t}^* \nabla^t_{\d^{r_t}_{\alpha_1}}\ldots f^*_{1/t} \nabla^t_{\d^{r_t}_{\alpha_n}}  W  \right|_{f_{1/t} (s^{t \mu}_c,t \mu)}.
\eeq
The point at which the transported covariant derivatives are evaluated approaches as $t\to \infty$ the fixed point representative $(s_*^\mu,\mu)$. Therefore, the renormalisation of correlation functions can be achieved by tuning  the renormalised operators and the renormalisation connection in a neighbourhood of the critical point in such a way that their transportation with $ f^*_t$ stays finite in the limit $t \to \infty$. That is, the renormalisation chart must be chosen such that the limits
\begin{align}
& \d^{r_*}_\alpha = \lim_{t\to \infty} f^*_{1/t} \d^{r_t}_\alpha, \\
& \nabla^{*} = \lim_{t\to \infty} f^*_{1/t} \nabla^t
\end{align}
are non-singular in a neighbourhood of the fixed point. Then, we get
\beq
G^R_{\alpha_1 \ldots \alpha_n}  =  \left. \nabla^*_{\d^{r_*}_{\alpha_1}}\ldots \nabla^*_{\d^{r_*}_{\alpha_n}}  W  \right|_{(s_*^{\mu},\mu)} .
\eeq
This simple but important result shows that the renormalised correlators are exactly equal to the bare correlators of the finite operators associated to $\d^{r_*}_{\alpha_i}$, evaluated at the fixed point. 
In arbitrary coordinates $c$, the components of the transported renormalised operators approach
\beq
\left.[\Ocal_{\alpha}^*]^{\alpha_1}\right|_{(s_*^{\mu},\mu)} = \lim_{t\to \infty}  M^{\alpha_1}_{t\,\alpha_2} \, \left.[\Ocal_{\alpha}^t]^{\alpha_2}\right|_{(s_c^{t\mu},t\mu)} , \label{linearren}
\eeq
where
\begin{align}
M^{\alpha_1}_{t\,\alpha_2} & =   \left. \d^c_{\alpha_2}  \right|_{(s_c^{t\mu},{t\mu})}  f^{\alpha_1}_{1/t}      \nn
& = \left. \frac{\delta f_{1/t}^{\alpha_1}[\gamma^{t\mu};g]}{\delta g^{\alpha_2}} \right|_{g_c^{t\mu}} 
\end{align}
is the Jacobian matrix of the coordinate transformation. Here, $g_c^\Lambda=c^\pi_\Lambda(s_c^\Lambda)$.\footnote{Note that $g_c^\Lambda$ and $g_c^{\Lambda^\prime}$ are related by a dilation, so $g_c^\Lambda$ will be independent of $\Lambda$ in the usual case of a homogeneous critical point.}
The components of the renormalised operators must be defined such that the limit in~\refeq{linearren} is finite and non-singular.\footnote{This may be impossible if the bare couplings $h_t$ are restricted to lie on a given ``bare'' submanifold, since in that case $[\Ocal_\alpha]^{\alpha_1}=0$ for directions $\alpha_1$ not parallel to that submanifold. In other words, we are using in this case a set of coordinate vectors $\{\d^c_\alpha\}$ that is not complete. As we will see in detail in \cite{LPVholography}, traditional holographic renormalisation with Dirichlet boundary conditions imposes such a restriction to single-trace operators and is in general insufficient to make all correlation functions well-defined.} This linear (or multiplicative, in the matrix sense) renormalisation can be found in the standard way, without knowledge of the RG flows, by requiring the cancellation of the non-local divergences in the correlation functions. But~\refeq{linearren} shows that a solution to this problem always exists:
\beq
\left.[\Ocal^t_\alpha]^{\alpha_1}\right|_{(s^{t\mu}_c,t\mu)}  = \left. \left(M_t^{-1} \right)^{\alpha}_{\alpha_1} \right|_{\mathrm{local}} ,\label{Mt}
\eeq
where ``local'' indicates a truncation of the derivative expansion that does not modify the limit in~\refeq{linearren}. This truncation is always possible since the derivatives come along with negative powers of $t$, which decrease the degree of divergence. With the choice in~\refeq{Mt}, the trivial limit $[\Ocal_\alpha^*]^{\alpha_1}=\delta^{\alpha_1}_\alpha$ is obtained.
Likewise, the transported Christoffel symbols approach
\beq
\left.\Gamma^{*\,\alpha_1}_{\alpha_2\alpha_3}\right|_{(s_*^{\mu},\mu)} =  \lim_{t\to \infty}  \left(M_t^{-1} \right)^{\alpha_5}_{\alpha_2} \left(M_t^{-1}\right)^{\alpha_6}_{\alpha_3}
\left( M^{\alpha_1}_{t\, \alpha_4} \left. \Gamma_{\alpha_5 \alpha_6}^{t\,\alpha_4}\right|_{(s_c^{t\mu},t\mu)} - \left.  \frac{\delta^2 f_{1/t}^{\alpha_1}[\gamma^{t\mu};g]}{\delta g^{\alpha_1} \delta g^{\alpha_2}} \right|_{g_c^{t\mu}} \right) 
\label{nonlinearren}
\eeq
in the limit. Observe that the transportation is non-linear, so a non-vanishing $\Gamma^t$ is necessary in general. An exception occurs for generic dimensions in normal coordinates, for which the second derivatives of the flows vanish. The renormalisation of the Christoffel symbols that keep $\Gamma^*$ finite can be obtained without knowledge of the RG flows by requiring the cancellation of local and semi-local divergences in the correlation functions. \refeq{nonlinearren} shows that, again, at least one solution exists:
\beq
\left. \Gamma^{t\, \alpha_1}_{\alpha_2\alpha_3}  \right|_{(s_c^{t\mu},t\mu) }= \left.\left( M_t^{-1} \right)^{\alpha_1}_{\alpha_4}  \left. \d_{\alpha_2} \d_{\alpha_3} f_{1/t}^{\alpha_4} \right|_{(s_c^{t\mu},t\mu) } \right|_{\mathrm{local}}. \label{Gammat}
\eeq
With this particular choice, $\left.\Gamma^{*\,\alpha_1}_{\alpha_2\alpha_3}\right|_{(s_c^{\mu},\mu)} = 0$. More generally,
we can always choose coordinates for which $\d_\alpha^{r_*}$ is a tangent vector at the fixed point. Furthermore, because the transportation preserves the flatness of the connection, it is possible to find, simultaneously, coordinates in which the Christoffel symbols vanish. In such coordinates $\tilde{c}$, the renormalised functions simply read
\beq
G^R_{\alpha_1 \ldots \alpha_n}  =  \left. \d^{\tilde{c}}_{\alpha_1} \ldots \d^{\tilde{c}}_{\alpha_n}  W  \right|_{(s_*^{\mu},\mu)} . \label{tildeR}
\eeq
If $c(s_*^{\mu},\mu)=0$, the coordinates $\tilde{c}$ are given, perturbatively, by
\beq
c^\alpha_{\mu} = \left.[\Ocal^*_{\alpha_1}]^\alpha\right|_{(s_*^{\mu},\mu)} \tilde{c}^{\alpha_1}_{\mu}
- \frac{1}{2} \left.\Gamma^{*\,\alpha}_{\alpha_3\alpha_4} [\Ocal^*_{\alpha_1}]^{\alpha_3}  [\Ocal^*_{\alpha_2}]^{\alpha_4}\right|_{(s_*^{\mu},\mu)} \tilde{c}^{\alpha_1}_{\mu}\tilde{c}^{\alpha_2}_{\mu} + O\left(\tilde{c}_{\mu}^3\right) . 
\eeq
In the renormalisation scheme given by~\refeq{Mt} and~\refeq{Gammat}, we directly have $\tilde{c}=c$.

In fact, the particular renormalised operators and Christoffel symbols in~\refeq{Mt} and~\refeq{Gammat} result, after trunctation, from the following choice of renormalisation coordinates (with the same $c$ as in those equations):
\beq
r^\alpha_{t}(s,\Lambda) = c^\alpha \circ f_{1/t}(s,\Lambda) - c^\alpha \circ f_{1/t}(s_c^{\Lambda},\Lambda) . \label{rf}
\eeq
The second term is included to ensure $r^\alpha_t(s_c^\Lambda,\Lambda)=0$.
This choice obviously gives finite correlation functions, as it fulfils~\refeq{rflimit}, with the very same $c$ used in the definition~\refeq{rf}. The result~\refeq{tildeR} with $\tilde{c}=c$ can then be understood as a direct consequence of this fact. Conversely, the renormalised operators and connection given by~\refeq{Mt} and~\refeq{Gammat} are the transportation with $f_t^*$ of the coordinate vectors $\d_\alpha^c$ and of the connection with vanishing Cristoffel symbols in $c$ coordinates. 

In particular, we can work in the renormalisation scheme where $r_t$ is given by~\refeq{rf} taking as $c$ a normal chart, $c=\bar{c}$. Such a renormalisation scheme will be called a {\em UV scheme} in the following.  UV schemes have the virtue of making the renormalisation operators and connection as simple as possible, when expressed in normal coordinates. In particular, when $s_c=s_*$,~\refeq{rf} can be inverted using $f_{1/t}^{-1}=f_t$:
\begin{equation}
\bar c^{\alpha}=f_t^{\alpha} [t^{-2}\bar \gamma;r_t^{\pi}].
\end{equation} 
Then, from~\refeq{generalct} and~\refeq{resonantflow} it follows that 
\begin{equation}
\left.[O_{\alpha^{\prime}}^t]^{\alpha}C^{t\,\alpha^{\prime}}_{\alpha_1\dots \alpha_n}\right|_{(s_*^{\Lambda},\Lambda)} =n!\,t^{-\left[\lambda_{(\alpha_1)}+\ldots+\lambda_{(\alpha_n)}\right]} \sum_{p=1}^{p^{\mathrm{max}}_{\alpha_1\cdots\alpha_n}}\log^p t \,\, [B_p]^{\alpha}_{\alpha_1\dots \alpha_n}\left(\gamma^{\Lambda}\right).
\end{equation} 
Therefore, the counterterms can be built in terms of beta coefficients, as illustrated in Fig.~\ref{fig:tree}. This structure agrees with Zimmermann's forest formula in perturbative coordinate-space renormalisation~\cite{Zimmermann:1969jj,Deser:1970spa}. Each forest for a given diagram is associated to a tree in~\refeq{tree}. Note in particular that in each forest only nested or disjoint (i.e. not overlapping) subtractions appear, which is an obvious property of~\refeq{tree} (see Fig.~\ref{fig:tree}). Moreover, the number of logs in a term corresponding to a given tree/forest is equal to the number of beta coefficients in the tree.

\subsection{Minimal subtraction}

Working in a UV scheme seems to require knowledge of the exact RG flows and their maximal diagonalisation. However, in this section we show that UV schemes are actually equivalent to certain minimal-subtraction schemes, defined in arbitrary coordinates without any explicit reference to RG flows or normal coordinates.  Our definition of minimal subtraction is given by the following restriction on the renormalisation chart:
\beq
r^\alpha_{t} = t^{\tilde{\lambda}_{(\alpha)}} R^\alpha_t\left[\bar \gamma;c^{\pi}\right],
\eeq
where $c$ is any chart, $\tilde \lambda_{(a x)} =\tilde \lambda_{(a)}$ are real numbers and $R_t^\alpha$ is a quasilocal functional of $g=c^{\pi}(s,\Lambda)$ that depends at most logarithmically on $t$ at each order in $c^\pi$. If this condition is met for a chart $c$, then it will also hold (with a different $R_t$) when $c$ is replaced by any other chart, in particular by a normal chart $\bar{c}$. Therefore, \refeq{rflimit} can be written as
\begin{align}
c^\alpha &  =  \lim_{t\to\infty} \left(r^\alpha_t \circ \bar{c}^{-1}\right) \circ \left(\bar{c} \circ f_t\right) \nn
         &  =  \lim_{t\to\infty} t^{\tilde{\lambda}_{(\alpha)}} \bar{R}^\alpha_t\left[t^2\bar \gamma;\bar{c}^{\pi} \circ f_{t} \right]. \label{minimalproof}
\end{align}
In the last line, the second entry in $R_t^\alpha$ is the RG flow in normal coordinates. As explained in Subsection~\ref{sec:normal}, \refeq{resonantflow}, the latter has coordinates of the form
\beq
\bar{c}^{\alpha} \circ f_{t}=\bar f^{\alpha}_{t} = t^{-\lambda_{(\alpha)}} F^\alpha_t\left[\bar\gamma;\bar{c}^{\pi}\right]=F^\alpha_t\left[t^2\bar\gamma;t^{-\lambda} \bar{c}^{\pi}\right], 
\eeq
where $\left(t^{-\lambda} \bar{c}^{\pi}\right)^{\alpha}=t^{-\lambda_{(\alpha)}} \bar{c}^{\alpha}$ and $F^\alpha_t$ is a quasilocal functional of $g^{\alpha}=c^{\alpha}(s,\Lambda)$ which can be expanded in a series of resonant monomials, with coefficients that depend at most logarithmically in $t$. Expanding the right-hand side of~\refeq{minimalproof} to linear order, we have
\beq
c^\alpha  = \lim_{t\to\infty} \left(A^\alpha_{\alpha_1} \bar{c}^{\alpha_1} t^{\tilde{\lambda}_{(\alpha)}-\lambda_{(\alpha_1)}} (1 +\mathrm{logs})+ O(\bar{c}^2) \right).
\eeq
But for the left-hand side of~\refeq{minimalproof} to be a non-singular invertible change of coordinates,  the linear term in its $\bar{c}$ expansion must be given by a finite, non-singular matrix multiplying the vector $\bar{c}$. This forces the matrix $A$ above to be non-singular and upper-triangular (if the directions are ordered with decreasing eigenvalues), and $\tilde{\lambda}_{(\alpha)}=\lambda_{(\alpha)}$. At the non-linear level, the terms with $\Pi_{i=1}^r \bar{c}_{\alpha_i}$, $n_{(\gamma)}$ metrics and $n_{(\gamma^{-1})}$ inverse metrics will scale like $t^{\lambda_{(\alpha)}-\sum_{i=1}^r \lambda_{(\alpha_i)}+ 2n_{\left(\gamma\right)}-2n_{(\gamma^{-1})}}$, up to logarithms. So, a finite continuum limit requires that each monomial either vanishes in the limit or is resonant (see~\refeq{resonance} and~\refeq{ConstraintNs2}).
Therefore, we learn that in minimal subtraction $c$ is a normal chart, just as in a UV scheme.

The restriction of minimal subtraction does not fix the renormalisation scheme completely. Indeed, if $r_t$ is a valid renormalisation chart in minimal subtraction, so is
\beq
\tilde r^{\alpha}_{t}= r_t^{\alpha}+ t^{\lambda_{(\alpha)}-\lambda_{(\alpha_1)}} a^{\alpha}_{t\,\alpha_1}(\bar \gamma)\, r_t^{\alpha_1}+ t^{\lambda_{(\alpha)}-\lambda_{(\alpha_1)}-\lambda_{(\alpha_2)}} a^{\alpha}_{t\,\alpha_1 \alpha_2}(\bar \gamma)\, r_t^{\alpha_1} r_t^{\alpha_2}+...\label{rsimplif} 
\eeq
if the coefficients $a^{\alpha}_{t\,\alpha_1\ldots \alpha_n}(\gamma)$ depend at most logarithmically on $t$ and vanish when $\sum_i \lambda_{(\alpha_i)}+2n_{\left(\gamma^{-1}\right)}-2n_{(\gamma)}\leq \lambda_{(\alpha)}$, being $n_{(\gamma)}$ and $n_{\left(\gamma^{-1}\right)}$ the number of metrics and inverse metrics respectively which $a^{\alpha}_{t\,\alpha_1\ldots \alpha_n}(\gamma)$ depends on. This ambiguity can be used to simplify $r_t$ and make it local.

In terms of renormalised operators and connections, minimal subtraction is characterised by the following two conditions:
\begin{description}
\item{(i)} The operator components $[\mathcal{O}^t_{\alpha_1}]^{\alpha_2}$ are required to be proportional to $t^{-\tilde \lambda_{\alpha_1}}$, up to logarithms, with the same non-negative number $\tilde \lambda_{\alpha_1}$ for all $\alpha_2$. 
\item{(ii)} The Christoffel symbols $\Gamma^{t\,\alpha}_{\alpha_1\alpha_2}$ are required to be $t$ independent in the neighbourhood where they are defined, up to logarithms.
\end{description}
The second condition can also be formulated in term of the counterterms $C^{t\,\alpha}_{\alpha_1...\alpha_n}$: they are required to be proportional to  $t^{\tilde \lambda_{\alpha}-\tilde \lambda_{\alpha_1}-...-\tilde \lambda_{\alpha_n}}$, up to logarithms, with the same $\tilde \lambda_{\alpha}$, $\tilde \lambda_{\alpha_1}$, ...,  $\tilde \lambda_{\alpha_n}$ as in condition (i). Again, there is some remaining freedom, which can be used to make the renormalised operators and counterterms local and to simplify them.


\section{Example: Gaussian fixed point}
\label{sec:free}

In this section we work out a simple example to illustrate the method and the general results found in the previous section. We consider the theory space of a single real scalar field $\omega$ in $d=4$ and examine the RG evolution close to the Gaussian fixed point. We compare with the renormalisation of composite operators in the free-field theory.

\subsection{Polchinski equation}

Following~\cite{Polchinski:1983gv}, we implement the cutoff procedure through a modified free propagator (of the dimensionless field),
\begin{align}
P_{\Lambda}^{xy} 
& = \langle \omega(x) \omega(y) \rangle_{(0,\Lambda)} \nn
& = P(\gamma^\Lambda;x-y) ,
\end{align}
where
\beq 
P(\gamma;x) = \frac{1}{4\pi^2} D \left(-\d^2_\gamma \right) \frac{1}{x^2_{\gamma}},  \label{propagator}
\eeq
with 
$\d^2_\gamma = \gamma^{\mu\nu} \d_\mu\d_\nu$, $x^2_\gamma = \gamma_{\mu\nu} x^\mu x^\nu$, and $D(u)$ a function with $D(0)$=1, decreasing sufficiently fast as $u\to \infty$. We also assume that $D(u)$ is analytic at $u=0$, which is required to keep the regularised actions quasilocal in fields and couplings. 

In a canonical linear parametrisation, the general Wilson action reads
\begin{align}
S[\gamma^\Lambda;g,\omega] = &\int d^4 x \sqrt{|\gamma^\Lambda|} \big[ g_0(x) 
+ g_1(x) \omega(x) + g_2(x) \omega(x)^2 + g_{2,2} \gamma^{\Lambda\, \mu \nu} \partial_{\mu}  \omega(x) \partial_{\nu}  \omega(x)    \nn 
& \mbox{} + g_3(x) \omega(x)^3 + \ldots \big],
\end{align} 
with the dots referring to the sum of couplings times other possible monomials in $\omega$ and its derivatives of arbitrary order, up to total derivatives. 
The partition function is given by
\begin{equation}
Z_{\Lambda}(s)=\frac{\int \mathcal{D}\omega  \exp\left\{-\int d^4x \sqrt{|\gamma^\Lambda|} \frac{1}{2} \gamma^{\Lambda \, \mu\nu} \d_\mu \omega(x) D^{-1}(-\d^2/\Lambda^2) \d_\nu \omega(x)  -s[\omega]\right\}}{\int \mathcal{D}\omega  \exp\left\{-\int d^4x \sqrt{|\gamma^\Lambda|} \frac{1}{2} \gamma^{\Lambda \, \mu\nu} \d_\mu \omega(x) D^{-1}(-\d^2/\Lambda^2) \d_\nu \omega(x) \right\}}.\label{PartitionFunctPol1}
\end{equation}
This equation can be understood as a specific implementation of~\refeq{partitionfunction}. The normalisation allows us to keep track of vacuum energy terms. Since we know the form of the cutoff, the Callan-Symanzik equation~\refeq{CallanSymanzik} can be made more explicit. Differentiation with respect to $\Lambda$ leads to Polchinski's equation~\cite{Polchinski:1983gv}, which in position space and using our geometric language reads
\begin{align}
\beta S_\omega -\omega^x \frac{\delta S_\omega}{\delta \omega^x}=\,\frac{1}{2}\dot P^{xy}(\bar \gamma) \frac{\delta^2 S_\omega}{\delta \omega^x \delta \omega^y}-\frac{1}{2}\dot P^{xy}(\bar \gamma) \frac{\delta S_\omega}{\delta \omega^x} \frac{\delta S_\omega}{\delta \omega^y},\label{PolEqGeo}
\end{align}
where
\begin{align}
\dot  P(\gamma)^{xy} &=   \dot P(\gamma;x-y) \nn
 & =  - \left(2+2\gamma \frac{\partial}{\partial \gamma} \right) P(\gamma;x-y) \nn
& = \frac{2}{\sqrt{|\gamma|}} D^\prime\left(-\d_{\gamma}^2\right) \delta(x-y).  \label{dotP}
\end{align}
Both sides of \refeq{PolEqGeo} are functions on $\Wcal$. This equation is satisfied also by the field-independent terms in the action, neglected in~\cite{Polchinski:1983gv}, when the cutoff dependence of the denominator in~\refeq{PartitionFunctPol1} is taken into account. Given a chart $c$, the first term of~\refeq{PolEqGeo} is written in components as
\begin{equation}
\beta S_\omega = \beta^{\alpha} \partial^c_{\alpha} S_\omega  +  2\bar \gamma \frac{\partial}{\partial \bar \gamma} S_\omega.
\end{equation} 
The point $s=0$ is trivially a fixed point. It corresponds to the free massless theory, since the kinetic term is included in the cutoff procedure.\footnote{There are strong indications that this is the only fixed point for the scalar field theory in four dimensions~\cite{Glimm:1987ng,Fernandez:1992jh}.}  In the following we study the theory space of one real scalar field in a neighbourhood of this Gaussian fixed point.

\subsection{Normal coordinates}
\label{sec:GausNormal}
To find the normal coordinates around the Gaussian fixed point and the corresponding beta functions, we could write 
\refeq{PolEqGeo} explicitly in an arbitrary parametrisation $c$, solve for the components $\beta^\alpha$ and finally find the change of coordinates that puts the betas in normal form. We will follow instead a more direct procedure in which we impose the normal form to~\refeq{PolEqGeo} (written in coordinates) from the very beginning and extract the normal beta functions and normal coordinates simultaneously. To do this, we first expand a general action close to the fixed point $s=0$ in normal coordinates:
\begin{equation}
S_\omega= (S^{\bar \gamma}_{\omega})_{\alpha}\, \bar c^{\,\alpha}+ (S^{\bar \gamma}_{\omega})_{\alpha_1 \alpha_2}\,\bar c^{\,\alpha_1} \bar c^{\,\alpha_2}+O(\bar c^{\,3}).  \label{normalS}
\end{equation}
Unlike $S_\omega$ and $c^\alpha$, which are non-trivial functions on $\Wcal$, the coefficients $(S^{\bar \gamma}_{\omega})_{\alpha_1\ldots \alpha_n}$ are functionals of $\omega$ that do not depend on the first component of the point in $\Wcal$, but only on the scale (as explicitly indicated by the $\bar \gamma$ superindex).  
The chart $\bar{c}$ is perturbatively defined by these coefficients. Next, we plug~\refeq{normalS} and \refeq{betacoordinates} (with $c=\bar{c}$) in~\refeq{PolEqGeo}), impose the normal condition~\refeq{resonantbeta} and solve order by order in $\bar{c}$ for the coefficients $\lambda^\alpha_{\alpha_1}$, $\bar \beta^\alpha_{\alpha_1 \ldots \alpha_n}$ and $(S^{\bar \gamma}_\omega)_{\alpha_1\ldots \alpha_n}$.

\subsubsection{Eigendirections}
We start with the linear order, assuming a diagonal\footnote{The fact that $\lambda$ is diagonalisable is an assumption that will be justified {\it a posteriori.}} matrix $\lambda^\alpha_{\alpha_1} = \lambda_{(\alpha)} \delta^\alpha_{\alpha_1}$ and $\lambda_{(ax)} = \lambda_{(a)} = 4- \Delta_{(a)}+n^u_{(a)}-n^d_{(a)}$, with $\Delta_{(a)}$ the conformal dimension of the operator and $n^u_{(a)}$ ($n^d_{(a)}$) the number of contravariant (covariant) indices of the coupling. At this order, \refeq{PolEqGeo} reduces to the following eigenvalue problem
\begin{equation}
\left[ \omega^x \frac{\delta}{\delta \omega^x} + \frac{1}{2}\dot P^{xy}(\gamma) \frac{\delta^2 }{\delta \omega^x \delta \omega^y} - 2 \gamma \frac{\d}{\d \gamma} \right] \left(S^{\gamma}_{\omega}\right)_{a}^z= \left(\Delta_{(a)}-n^u_{(a)}+n^d_{(a)} \right)\left(S^{\gamma}_{\omega}\right)_{a}^z . \label{EigenDirEq}
\end{equation}
Recall that, in our covariant notation,  $(S^{\gamma}_{\omega})_{ax}=\sqrt{|\gamma|}(S^{\gamma}_{\omega})_{a}^x$, with $(S^{\gamma}_{\omega})_{a}^x$ a scalar.  This calculation of eigenoperators of the Polchinski equation in a scalar theory has been performed before in~\cite{Rosten:2010vm}. 
A trivial solution is given by the identity operator,
\beq
\left(S_\omega^\gamma\right)_0^x = 1 ,
\eeq
with dimension $\Delta_{(0)} = 0$.\footnote{Note that the unitarity bound $\Delta \geq 1$ does not apply to this field-independent operator.} To find the non-trivial solutions,
we make an Ansatz consistent with our requirement of a quasilocal Wilson action:
\begin{equation}
\left(S^{\gamma}_{\omega}\right)_{a}^x=\mathcal{Q}_{a,x_1...x_{m_a}}^{x}(\gamma) \left(S^{\gamma}_{\omega}\right)^{\langle x_1...x_{m_a}\rangle},\label{EigenDirForm}
\end{equation} 
where $m_a$ is a positive integer associated to $a$ and $\mathcal{Q}_{a,x_1...x_n}^{x}(\gamma)$ is a product of Dirac deltas and their derivatives with support at $x=x_1=...=x_n$, while $\left(S^{\gamma}_{\omega}\right)^{\langle x_1...x_n\rangle}$ are functions of $x_1,...,x_n$ and of $\omega(x_1),...,\omega(x_n)$, with the requirement that they and their derivatives of any order are well defined at coincident points.
Inserting this form of the eigenfunctions into~\refeq{EigenDirEq}, we find that the $\left(S^{\gamma}_{\omega}\right)^{\langle x_1...x_{m_a}\rangle}$ must be solutions of the equation
\begin{equation}
\left[\Delta_{(a)} - n^{(a)}_\d+2 \gamma \frac{\partial}{\partial \gamma} -\omega^x \frac{\delta}{\delta \omega^x}-\frac{1}{2}\dot P^{xy}(\gamma) \frac{\delta^2 }{\delta \omega^x \delta \omega^y}\right]\left(S^{\gamma}_{\omega}\right)^{\langle x_1...x_{m_a}\rangle}=0,\label{EigenMultiEq}
\end{equation} 
with $n^{(a)}_\d$ the number of derivatives in the corresponding $\mathcal{Q}$. The existence of a solution $\left(S^{\gamma}_{\omega}\right)^{\langle x_1...x_{m_a}\rangle}$ to this equation requires $\Delta_{(a)} = n^{(a)}_\d + m_a$, which gives rise to the discrete spectrum $\Delta_{(a)} \in \mathbb{N}$ (besides $\Delta_{(0)}=0$ for the identity). The first four explicit solutions are
\begin{align}
&\left(S^{\gamma}_{\omega}\right)^{\langle x\rangle}=\omega^{x_1},\nn
&\left(S^{\gamma}_{\omega}\right)^{\langle x_1x_2\rangle}= \omega^{x_1}  \omega^{x_2}-P^{x_1x_2}(\gamma),\nn
& \left(S^{\gamma}_{\omega}\right)^{\langle x_1x_2x_3\rangle}=\omega^{x_1} \omega^{x_2} \omega^{x_3}-\left[\omega^{x_3} P^{x_1x_2}(\gamma) + \omega^{x_2} P^{x_1x_3}(\gamma)+\omega^{x_1}P^{x_2x_3}(\gamma)  \right],\nn
& \left(S^{\gamma}_{\omega}\right)^{\langle x_1x_2x_3x_4\rangle}=\omega^{x_1} \omega^{x_2} \omega^{x_3} \omega^{x_4}-\left[ \omega^{x_1} \omega^{x_2}P^{x_3x_4}(\gamma) +  \mathrm{5~inequivalent~permutations}  \right]\nn
&\hspace{1.8cm}+\left[P^{x_1x_2}(\gamma)P^{x_3x_4}(\gamma) +P^{x_1x_3}(\gamma)P^{x_2x_4}(\gamma)+P^{x_1x_4}(\gamma)P^{x_2x_3}(\gamma) \right].\label{EigenPert}
\end{align}
These functionals have a remarkable property: their expectation value vanishes in the free theory,
\begin{align}
 &\left \langle \left(S^{\gamma^{\Lambda}}_{\omega}\right)^{\langle x_1... x_n \rangle} \right\rangle_{(0,\Lambda)}=0,\label{ExpValEig0}
\end{align}
for $n\geq 1$. For odd $n$ this statement is trivial, while for even $n$ it can be checked contracting the fields $w^x$ in~\refeq{EigenPert} with Wick's theorem.

The local eigenoperators $\left(S^{\gamma}_{\omega}\right)_{a}^x$  can be constructed from these solutions using~\refeq{EigenDirForm}. For instance,
\begin{align}
& (S_\omega^\gamma)^x_1 = \omega(x) , \nn
& (S_\omega^\gamma)^x_2 = \left( \omega(x) \right)^2 - P(\gamma;0) , \nn
& (S_\omega^\gamma)^x_{2_2} = \omega(x)  \d_\gamma^2 \omega(x) -  \d_\gamma^2 P(\gamma;0) , \nn
& (S_\omega^\gamma)^x_{3} = \left(\omega(x)\right)^3   - 3 P(\gamma;0) \omega(x) ,  \label{explicitS}
\end{align}
where $1,2,2_2,3,\ldots$ label the eigendirections. Their eigendimensions are, respectively, $\Delta=1,2,4,3$, in agreement with the conformal dimensions of a free theory. Note that $P(\gamma;0)$ and $\d_\gamma^n P(\gamma;0)$ for any $n$ are dimensionless constants. It is thus clear that any linear product of fields and their derivatives at $x$ can be written as a linear combination of the operators $\left(S^{\gamma}_{\omega}\right)_{a}^x$ constructed in this way, together with the identity $\left(S^{\gamma}_{\omega}\right)_{0}^x$. Therefore, these operators form a complete set and the linear part of the beta function is indeed diagonalisable.  
In the following we call $P_0=P(\gamma;0)$.

\subsubsection{Higher orders}

The higher orders can be obtained using~\refeq{PolEqGeo} iteratively. The quadratic term $(S^{\gamma}_{\omega})_{a_1x_1\,a_2x_2}=|\gamma|(S^{\gamma}_{\omega})^{x_1 x_2}_{a_1a_2}$ is given by
\begin{align}
& \left[2d-\lambda_{(a_1)}-\lambda_{(a_2)}+2 \gamma\frac{\partial}{\partial \gamma} -\omega^x \frac{\delta}{\delta \omega^x}-\frac{1}{2}\dot P^{xy}(\gamma) \frac{\delta^2 }{\delta \omega^x \delta \omega^y}\right] (S^{\gamma}_{\omega})^{z_1 z_2}_{a_1a_2} \nn
 & \mbox{} +\frac{1}{|\gamma|}\bar \beta^{\alpha}_{a_1z_1\, a_2z_2} (S^{\gamma}_{\omega})_{\alpha} =-\frac{1}{2}\dot P^{xy} \frac{\delta (S^{\gamma}_{\omega})^{z_1}_{a_1}}{\delta \omega^x}\frac{\delta (S^{\gamma}_{\omega})^{z_2}_{a_2}}{\delta \omega^y}.
\label{HOEq}
\end{align} 
The quadratic beta coefficients on the left hand side of the equation (second line) is required to be resonant. They are necessary to cancel possible non-localities. As an example, let us solve the equation for the action coefficient $(S^{\gamma}_{\omega})_{33}^{x_1x_2}$, where the eigendirection $a=3$ is defined in~\refeq{explicitS} and has $\lambda_{(3)}=1$. Choosing $a_1=a_2=3$,~\refeq{HOEq} reads
\begin{align}
&\left[6+2 \gamma \frac{\partial}{\partial \gamma} -\omega^x \frac{\delta}{\delta \omega^x}-\frac{1}{2}\dot P^{xy}(\gamma) \frac{\delta^2 }{\delta \omega^x \delta \omega^y}\right]  
(S^{\gamma}_{\omega})_{33}^{z_1z_2}+\frac{1}{|\gamma|}\left[ \bar \beta^{2z}_{3z_13z_2}(S^{\gamma}_{\omega})_{2z} + \bar \beta^{0z}_{3z_13z_2}(S^{\gamma}_{\omega})_{0z}\right]&\nn
& =-\frac{9}{2} \dot P(\gamma,z_1-z_2)\left[(S^{\gamma}_{\omega})^{\langle z_1z_1 z_2z_2 \rangle}+4P^{z_1z_2}(\gamma)(S^{\gamma}_{\omega})^{\langle z_1z_2 \rangle}  +2P^{z_1z_2}(\gamma)^2\right].\label{PolEqS33}
\end{align} 
We have included the only possible resonant beta terms:
\begin{align}
\bar \beta^{2z}_{3z_13z_2}(\gamma) &= b^{2}_{33} \delta(z-z_1)\delta(z-z_2) , \\
\bar \beta^{0z}_{3z_13z_2}(\gamma) &= \frac{b^{0}_{33}}{2} \left[\delta(z-z_1)\partial^2_{\gamma}\delta(z-z_2)+\partial^2_{\gamma}\delta(z-z_1)\delta(z-z_2)\right] .
\end{align}
The values of the coefficients $b^{0,2}_{33}$ will be determined below.
To solve~\refeq{PolEqS33}, let us make the ansatz
\begin{align}
(S^{\gamma}_{\omega})_{33}^{xy}&=A(\gamma;x-y)\, (S^{\gamma}_{\omega})^{\langle xxyy \rangle}+B(\gamma;x-y) (S^{\gamma}_{\omega})^{\langle xy \rangle}+E(\gamma;x-y).\label{S33}
\end{align}
Using~\refeq{EigenMultiEq},
we find that the functions $A$, $B$ and $E$ must satisfy
\begin{align}
&\left[2+2 \gamma_{\mu \nu}\frac{\partial}{\partial \gamma_{\mu \nu}}\right] A(\gamma;x)=-\frac{9}{2} \dot P(\gamma;x),\label{AEq} \\
&\left[4+2 \gamma_{\mu \nu}\frac{\partial}{\partial \gamma_{\mu \nu}}\right] B(\gamma;x)=-18 \dot P(\gamma;x)P(\gamma;x)-b^{2}_{33}\frac{\delta(x)}{\sqrt{|\gamma|}}, \label{BEq} \\
&\left[6+2 \gamma_{\mu \nu}\frac{\partial}{\partial \gamma_{\mu \nu}}\right] E(\gamma;x)=-9 \dot P(\gamma;x)P(\gamma;x)^2-b^{0}_{33}\frac{\partial_{\gamma}^2\delta(x)}{\sqrt{|\gamma|}}.\label{EEq}
\end{align}
The most general solutions of these three equations are
\begin{align}
A(\gamma;x)&=\frac{9}{2}P(\gamma;x)-\frac{\xi_1}{4\pi^2 x_{\gamma}^2}, \label{Asol} \\
B(\gamma;x)&=9P(\gamma;x)^2+\frac{b^2_{33}}{8\pi^2} \dot \partial^2_{\gamma}\left[ \frac{\log x_{\gamma}^2}{x_{\gamma}^2}\right]+\xi_2\frac{\delta(x)}{\sqrt{|\gamma|}} ,\label{Bsol} \\
E(\gamma;x)&=3P(\gamma;x)^3+\frac{b^0_{33}}{8\pi^2} \,\dot \partial^4_{\gamma}\left[ \frac{\log x_{\gamma}^2}{x_{\gamma}^2}\right] +\xi_3 \frac{\partial_{\gamma}^2\delta(x)}{\sqrt{|\gamma|}}, \label{Esol}
\end{align}
where 
the arbitrary parameters $\xi_1$, $\xi_2$ and $\xi_3$ are associated to solutions to the homogeneous part of~\refeq{AEq}, \refeq{BEq} and~\refeq{EEq}. The dot on the derivatives $\dot \partial^2$ and $\dot \partial^4$ indicates that they are defined in the sense of distributions, acting by parts on test functions and discarding (singular) surface terms. Then, $A(\gamma;x)$, $B(\gamma;x)$ and $E(\gamma;x)$ are well-defined distributions.\footnote{Acting instead with the derivatives on the functions inside the brackets results in 
\begin{align}
 &\partial^2_{\gamma}\left[ \frac{\log x_{\gamma}^2}{x_{\gamma}^2}\right]=-\frac{4}{x_{\gamma}^4},\\
 &\partial^4_{\gamma}\left[ \frac{\log x_{\gamma}^2}{x_{\gamma}^2}\right]=-\frac{32}{x_{\gamma}^6},
\end{align}
which are too singular at $x=0$ to admit a Fourier transform. The expressions with the dotted derivatives correspond to the renormalised values of these functions in differential renormalisation~\cite{Freedman:1991tk}.
}
Their asymptotic behaviour when $x\to\infty$ is given by
\begin{align}
A(\gamma;x)&\sim \left( \frac{9}{2} -\xi_1\right) \frac{1}{4\pi^2 x_{\gamma}^2} , \\
B(\gamma;x)&\sim \left( 9 -8\pi^2 b^{2}_{33}\right) \frac{1}{\left(4\pi^2\right)^2 x_{\gamma}^4} , \\
E(\gamma;x)&\sim \left( 3 -256\pi^4 b^{0}_{33}\right) \frac{1}{\left(4\pi^2\right)^3 x_{\gamma}^6} . \label{AsymptBehx}
\end{align} 
This shows that these functions are in general non-local, with Fourier transforms that behave like $p^{-2}$, $\log p^2$ and $p^2 \log p^2$ as $p \to 0$ and are thus non-analytic at $p=0$. To ensure that $(S^{\gamma}_{\omega})_{33}^{z_1z_2}$ is quasilocal we need to fix $\xi_1 = 9/2$, $b_{33}^2=9/(8\pi^2)$ and $b_{33}^0 = 3/(256 \pi^4)$. Therefore, the
beta functions take the values
\begin{align}
\bar \beta^{2z}_{3x,3y}(\gamma) &=\frac{9}{8\pi^2}\delta(z-x)\delta(z-y),\label{beta2} \\
\bar \beta^{0z}_{3x,3y}(\gamma) &=\frac{3}{512\pi^4}\left[\delta(z-x)\partial^2_{\gamma}\delta(z-y)+\partial^2_{\gamma}\delta(z-x)\delta(z-y)\right]. \label{beta0}
\end{align}
We stress that they do not depend on the particular regulator $P$. The beta function $\bar \beta^{0z}_{3x,3y}$ is associated to the identity operator and represents a contribution to the conformal anomaly in the presence of local couplings, see~\refeq{splitCallanSymanzik}:
\begin{equation}
\mathcal{A}(x)=\frac{3}{256\pi^4} \bar c^3(x) \partial^2_{\gamma} \bar c^3(x) + \ldots
\end{equation}
As long as the regulating function $D(u)$ is analytic at $u=0$, as assumed, the functions $A$, $B$ and $E$ with the selected values of $\xi_1$, $b_{33}^2$ and $b_{33}^0$ can be expanded as a series of local distributions,
\begin{align}
A(\gamma;x-y)&=\frac{1}{\sqrt{|\gamma|}}\left[A_0\, \delta(x-y) + A_2\, \partial_{\gamma}^2 \delta(x-y) +A_4\, \partial_{\gamma}^4 \delta(x-y)+ ...  \right],\nn
B(\gamma;x-y)&=\frac{1}{\sqrt{|\gamma|}}\left[B_0\, \delta(x-y) + B_2\, \partial_{\gamma}^2 \delta(x-y) +B_4\, \partial_{\gamma}^4 \delta(x-y)+ ...  \right],\nn
E(\gamma;x-y)&=\frac{1}{\sqrt{|\gamma|}}\left[E_0\, \delta(x-y) + E_2\, \partial_{\gamma}^2 \delta(x-y) +E_4\, \partial_{\gamma}^4 \delta(x-y)+ ...  \right],\label{ABEExpansion}
\end{align}
with coefficients $A_n$, $B_n$ and $E_n$ depending on the chosen function $D$. Only $B_0$ and $E_2$ remain arbitrary, since they depend on the parameters $\xi_2$ and $\xi_3$. Using these functions in \refeq{EigenDirForm} we find $(S^{\gamma}_{\omega})_{33}^{xy}$, which  gives a quasilocal contribution to $s$ in~\refeq{normalS}. 


\subsection{Renormalised correlators}

The Wilsonian information provided by beta functions and normal coordinates allows us to calculate renormalised correlation functions with the methods of Section~\ref{sec:Correlation}. These calculations can be performed in two related ways. The most direct one is to take advantage of~\refeq{tildeR} to compute the renormalised correlators directly as bare correlators at the fixed point with a finite cutoff. This calculation is very simple when the fixed point is well-characterised---as is the free theory---since no counterterms are required and the continuum limit is not explicitly taken. Alternatively, we can use the RG flows to compute the corresponding renormalised operators and counterterms and then use \refeq{TwoPointFunct} to obtain the renormalised correlator. We will consider both approaches in turn.

\subsubsection{Normal correlators}

The bare correlation functions in normal coordinates are defined by functional derivatives of the generator $W$ with respect to normal couplings. We are interested in functional derivatives at the fixed point. Let us calculate
\begin{equation}
G^R_{3x 3y} = \left.\partial^{\bar c}_{3x} \partial^{\bar c}_{3y} W\right|_{(0,\mu)}.
\end{equation} 
The notation $G^R$ already anticipates that, as shown in~\refeq{tildeR}, this correlator computed at a finite cutoff $\mu$ is equal to the corresponding renormalised correlator, in the UV scheme given by~\refeq{rf} with $c=\bar{c}$.
From~\refeq{normalS}, using~\refeq{ExpValEig0},
\begin{equation}
\left.\partial^{\bar c}_{3x} \partial^{\bar c}_{3y} W\right|_{(0,\mu)}=
\left\langle \left(S^{\gamma^{\mu}}_{\omega}\right)_{3x} \left(S^{\gamma^{\mu}}_{\omega}\right)_{3y} \right\rangle_{(0,\mu)}
-2\left\langle \left(S^{\gamma^{\mu}}_{\omega}\right)_{3x3y} \right\rangle_{(0,\mu)}.
\end{equation}
\begin{figure}[t!]
\begin{center}
\includegraphics[width=12cm]{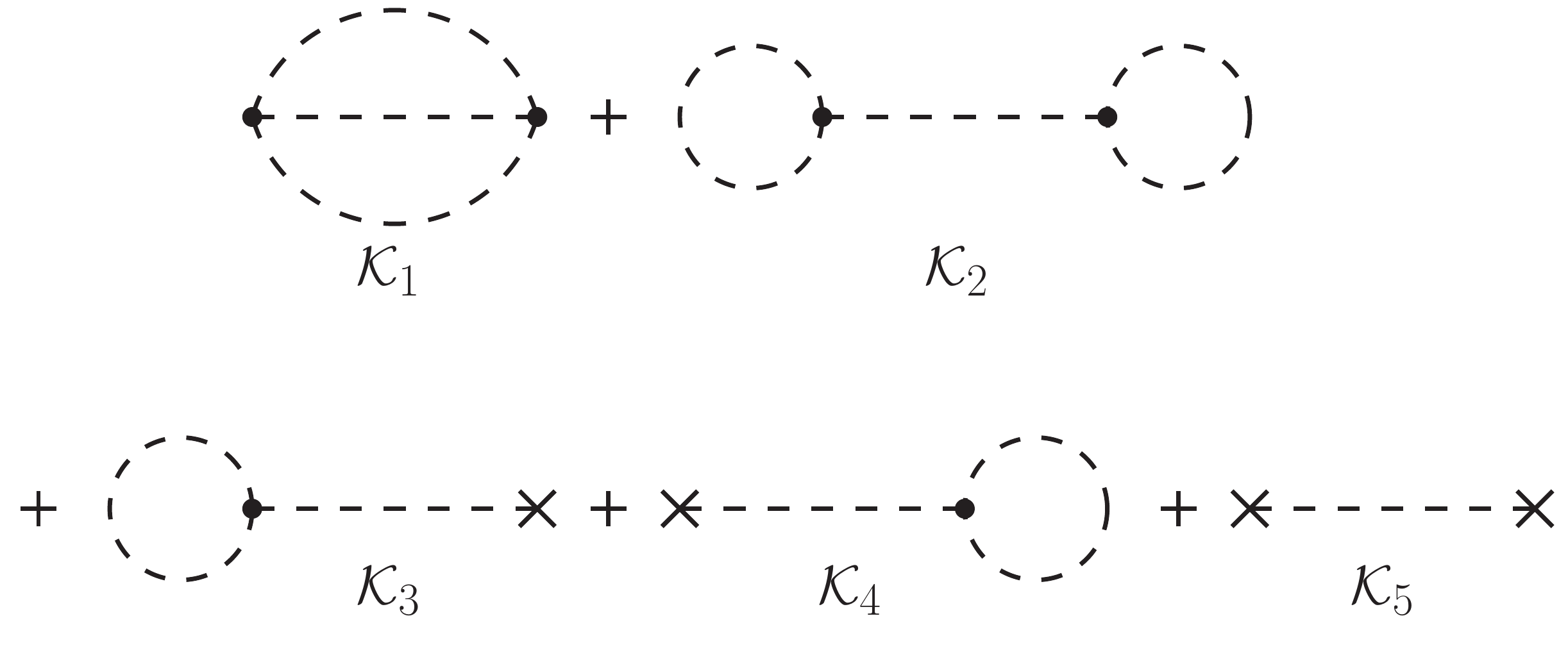}
\caption{Calculation of $\left\langle \left(S^{\gamma^{\mu}}_{\omega}\right)_{3x} \left(S^{\gamma^{\mu}}_{\omega}\right)_{3y} \right\rangle_{(0,\mu)}$. Lines indicate propagators $P(\gamma^\mu;x-y)$, dots represent vertices $\sqrt{|\gamma^{\mu}|}\,\omega(x)^3$, and crosses, insertions of $-3\sqrt{|\gamma^{\mu}|}P_0\,\omega(x)$. }
\label{fig:S3S3}
\end{center}
\end{figure}
Using Wick's theorem and the explicit form in~\refeq{explicitS},
we find that the first term on the left-hand side has contributions given by the diagrams $\mathcal{K}_1$--$\mathcal{K}_5$ in Fig.~\ref{fig:S3S3} in the free-field theory. 
Taking combinatorial factors into account, it is easy to check that $\sum_{i=2}^5 \mathcal{K}=0$. Therefore,
\begin{equation}
\left\langle \left(S^{\gamma^{\mu}}_{\omega}\right)_{3x} \left(S^{\gamma^{\mu}}_{\omega}\right)_{3y} \right\rangle_{(0,\mu)}
=6|\gamma^{\mu}|P(\gamma^{\mu};x-y)^3.
\end{equation} 
For the second term, using the explicit form~\refeq{S33} and the property~\refeq{ExpValEig0}, we get
\begin{equation}
\left\langle \left(S^{\gamma^{\mu}}_{\omega}\right)_{3x3y} \right\rangle_{(0,\mu)}=|\gamma^{\mu}|E(\gamma^{\mu};x-y).
\end{equation} 
Combining both results and inserting the solution~\refeq{Esol}, we finally obtain
\begin{equation}
\left.\partial^{\bar c}_{3x} \partial^{\bar c}_{3y} W\right|_{(0,\mu)}=-\frac{3\mu^4}{2^{10}\pi^6} \,\dot \partial^4\left[ \frac{\log (x-y)_{\gamma^\mu}^2}{(x-y)_{\gamma^\mu}^2}\right] -2\xi_3\mu^2 \partial^2\delta(x-y).  \label{reg33}
\end{equation} 
Note that the $P^3$ terms have cancelled out.
The arbitrary parameter $\xi_3$ multiplies a scheme-dependent local term. It can be absorbed into a redefinition of the scale $\mu$.

\subsubsection{Renormalisation}
\label{sec:WilRen}

The standard calculation of the renormalised correlators involves a renormalisation procedure: first finding universal cutoff-dependent renormalised operators and counterterms and then taking the continuum limit for the correlators of interest. Such a renormalisation can be carried out in arbitrary coordinates. Usually, the renormalised operators and counterterms are determined by requiring the corresponding contributions to cancel the continuum-limit divergences of the bare correlators. Here we show how to obtain them from the exact RG flows near the fixed point, using the results in \ref{sec:Correlation}. We concentrate on the renormalised operators and counterterms that contribute to $G^R_{33}$.

We choose a renormalisation chart given by~\refeq{rf} with normal coordinates $c=\bar{c}$, i.e.\ we choose a UV scheme. Since our fixed point is Gaussian, the simplest choice for the critical point is clearly the fixed point itself, $s_c = s_*=0$.\footnote{The relation with the calculation in the previous section is quite straightforward in this case, for the renormalisation chart simply counteracts the action of the RG flow. Indeed,~\refeq{rf} with $s_c=s_*$ and $c=\bar{c}$ amounts to the choice $h_t = f_t \circ \bar{c}^{-1}$ for the bare couplings, which used in~\refeq{correlationRG} directly gives normal correlators.} Using~\refeq{resonantflow} in~\refeq{rf} (with $c=\bar{c}$) we find an explicit relation between $\bar{c}$ and $r_t$. Inverting this relation perturbatively, we can write~\refeq{normalS} in terms of renormalisation coordinates:
\begin{equation}
S_{\omega}=r_t^{\alpha}\, t^{-\lambda_{(\alpha)}} \,(S^{\bar \gamma}_{\omega})_{\alpha}+r_t^{\alpha_1} r_t^{\alpha_2}\,t^{-\lambda_{(\alpha_1)}-\lambda_{(\alpha_2)}}\,\left[\log t \, \beta^{\alpha}_{\alpha_1\alpha_2}(\bar \gamma)\, (S^{\bar \gamma}_{\omega})_{\alpha}+ (S^{\bar \gamma}_{\omega})_{\alpha_1\alpha_2}   \right] + O\left(r_t^3\right).\label{swrt}
\end{equation} 
In the UV scheme, the renormalised operator associated to the eigendirection $3$ is given by
\beq
\left. \partial ^{r_t}_{3x} \right|_{(0,t\mu)} = f_{1/t}^* \left.\partial^{\bar c}_{3x}\right|_{(0,\mu)} .
\eeq
Since the renormalisation procedure is done around the critical point, all functions, vector and tensor fields on $\mathcal{W}$ in the formulas below are understood to be evaluated at the point $(0,t\mu)$, unless otherwise indicated.
We can directly read the renormalised operator from~\refeq{swrt}:
\begin{align}
\left[\mathcal{O}^t_{3x} \right] & = \partial ^{r_t}_{3x}\,S_{\omega} \nn
& = t^{-1} \left( S_\omega^{\gamma^{t \mu}}\right)_{3x} \nn
& =t^{-1}\sqrt{|\gamma^{t\mu}|}\left[ \omega(x)^3-P_{0}\, \omega(x)  \right].\label{RenOp3}
\end{align} 
In order to compare with standard perturbative calculations, we present the calculation in terms of canonical linear coordinates which are called $c$ hereafter. Using this parametrisation we can write~\refeq{RenOp3} as
\beq
\partial ^{r_t}_{3x}  = t^{-1} \left[\partial_{3x}^c-P_{0}\partial^c_{1x} \right].
\eeq
So, in this basis the non-trivial components of $\left. \partial ^{r_t}_{3x} \right|_{(0,t\mu)}$ are 
\begin{align}
& \left[ \Ocal_{3x}^t \right]^{3y} = t^{-1} \delta(x-y),  \label{Ot3} \\
& \left[ \Ocal_{3x}^t \right]^{1y} = - t^{-1} P_{0}\, \delta(x-y). \label{Ot1}
\end{align}
We will also need below
\begin{align}
\left[ \Ocal^t_{0x} \right] & =  \d_{0x}^{r_t} \, S_\omega \nn
& = t^{-4} \sqrt{|\gamma^{t\mu}|}  \nn
& = \mu^4 ,
\end{align}
which can be read from~\refeq{swrt} as well. In the canonical linear basis, its only non-vanishing component is
\beq
\left[ \Ocal_{0x}^t \right]^{0y} = t^{-4} \delta(x-y). \label{O0}
\eeq
We see that the vacuum energy operator, having the lowest dimension, does not mix with other operators. 

The nonlinear counterterms are defined by the transported connection $f_t^* \nabla$ in normal coordinates. They can be obtained from the non-linear part of~\refeq{swrt}, as we now show.
We make use of the relation
\beq
C^{t \, \alpha}_{\alpha_1 \alpha_2} \d_\alpha^{r_t} S_\omega =  \d^{r_t}_{\alpha_1} \d^{r_t}_{\alpha_2} S_\omega ~~~~~~~\mbox{(linear $c$)},
\label{linearC}
\eeq
which is valid for any linear parametrisation $c$ and follows from~\refeq{QuadCounterT} and the fact that in linear coordinates $\d^c_{\alpha_1} \d^c_{\alpha_2} S_\omega =0$. Choosing $\alpha_1=3x$, $\alpha_2=3y$ and taking~\refeq{RenOp3} and~\refeq{swrt} into account, we get the equation
\beq
C^{t\,\alpha}_{3x3y} \,t^{-\lambda_{(\alpha)}} (S^{\gamma^{t\mu}}_{\omega})_{\alpha}
=2t^{-2}\left[ \log t \, \beta^{\alpha}_{3x3y}(\gamma^{t\mu})\, (S^{\gamma^{t\mu}}_{\omega})_{\alpha}+ (S^{\gamma^{t\mu}}_{\omega})_{3x3y}  \right].\label{ExtractCT}
\eeq
To solve it for $C^{t\,\alpha}_{3x3y}$, we use~\refeq{S33} and~\refeq{ABEExpansion} to expand the last term inside the brackets in the basis of eigenoperators,
\begin{equation}
(S^{\gamma}_{\omega})_{3x3y} =  \left(\tau^{\gamma}\right)^{\alpha}_{3x 3y}  (S^{\gamma}_{\omega})_{\alpha}.
\end{equation} 
Note that this expansion includes not only scalar but also tensorial eigenoperators $(S^{\gamma}_{\omega})_{\alpha}$ in which the collective index $\alpha$ contains Lorentz indices (contracted with the ones in the coefficients). For our purposes we can cut the series and keep only the terms that will eventually contribute in the $t \to \infty$ limit, which in this case involve only scalar operators. This is equivalent to using~\refeq{rsimplif} to redefine the renormalisation scheme, with no impact in the final renormalised correlators. We find
\begin{align}
(S^{\gamma}_{\omega})_{3x3y} = & E_0  \sqrt{|\gamma|} \delta(x-y)  + E_2 \sqrt{|\gamma|} \d_\gamma^2 \delta(x-y) +  B_0 \delta(x-y) \left(S_\omega^\gamma\right)_{2x}    \nn
& + \mbox{irrelevant terms}.
\end{align}
Matching in~\refeq{ExtractCT} the coefficients of the identity and $\left(S_\omega^\gamma\right)^x_2 $ and using the values~\refeq{beta2} and \refeq{beta0} for the beta functions, we identify the counterterms
\begin{align}
C^{t\,2 z}_{3x3y}=&  \left[\frac{9}{4\pi^2} \log t  +2B_0 \right]\delta(z-x)\delta(z-y),\label{CTC2}\\
C^{t\,0 z}_{3x3y}=
&\left[\frac{3}{256 \pi^4} \log t +E_2 \right] \left[\delta(z-x)\partial_{\gamma^{\mu}}^2 \delta(z-y)+ \delta(z-y)\partial^2_{\gamma^{\mu}} \delta(z-x)\right]\nn
& +2t^{2}E_0\delta(z-x)\delta(z-y).\label{CTC0}
\end{align} 

Equipped with the renormalised operators and counterterms in linear coordinates, we are ready to compute the renormalised correlator $G^R_{3x3y}$ using~\refeq{TwoPointFunct} and~\refeq{NonLinCounterterms}: 
\begin{align}
G^R_{3x3y}&=\lim_{t\to\infty}\left\{  [\mathcal{O}^t_{3x}]^{\alpha_{1}} [\mathcal{O}^t_{3y}]^{\alpha_{2}} 
 \d^c_{\alpha_{1}} \d^c_{\alpha_{2}}  W +C^{t\,\alpha_1}_{3x3y} [\mathcal{O}^t_{\alpha_1}]^{\alpha_2}\partial^c_{\alpha_2} W  \right\}. 
\label{GR33}
\end{align}
This limit is well-defined for any valid cutoff propagator $P$, but to make the cancellation of divergences manifest let us choose the following simple cutoff propagator:
\begin{equation}
P(\gamma;x)=\frac{1}{4\pi^2}\frac{1}{x_{\gamma}^2+1}. \label{regprop}
\end{equation}
This corresponds to the function
\begin{align}
D(u) &= \sqrt{u} K_1(\sqrt{u}) \nn
& = 1+\frac{1}{4}\left( 2\gamma_{E}-1+\log \frac{u}{2} \right) u+o(u).
\end{align}
Note that it does not satisfy the requirement of analiticity at $u=0$. Even if this results in a non-quasilocal Wilson action, the non-local pieces are irrelevant for the continuum limit and we find the same renormalised propagator that could be obtained with an analytic (but more complicated) regularisation.\footnote{With the propagator~\refeq{regprop} the expansion of the functions $A,B,E$ needs to be modified to include non-local terms. In Fourier space,
\begin{align}
\hat A(\gamma;p)&=\frac{1}{\sqrt{|\gamma|}}\frac{9}{2}\left[\frac{2\gamma_E-1}{4}+\frac{1}{4}\log\frac{p_{\gamma}^2}{4}+ o\left(p_{\gamma}^0\right)  \right],\nn
\hat B(\gamma;p)&=\frac{1}{\sqrt{|\gamma|}}\left[B_0+\frac{9\,p_{\gamma}^2}{8 \pi^2}\left(-\gamma_E+\frac{1-\gamma_E}{4}-\frac{\log p^2_{\gamma}}{4}  \right)+o\left(p_{\gamma}^2\right)\right],\nn
\hat E(\gamma;p)&=\frac{1}{\sqrt{|\gamma|}}\left[\frac{3}{128 \pi^4}-p_{\gamma}^2\,E_2+ \frac{3p_{\gamma}^4}{2^{12}\pi^4} \left( \frac{4\gamma_E-5}{2}+\log \frac{p_{\gamma}^2}{4}\right)+ o\left(p_{\gamma}^4\right) \right].
\end{align}
Only the first terms, which are local, contribute in the $t\to \infty$ limit and we obtain the same counterterms as above, with $E_0=\frac{3}{128\pi^4}$.}

Using~\refeq{Ot3} and~\refeq{Ot1} and Wick's theorem, the first term on the right-hand side of~\refeq{GR33} is found to be given precisely by the diagrams in Fig.~\ref{fig:S3S3}, up to a global $t^{-2}$ factor and with $\mu$ changed by $t \mu$. Therefore,
\begin{align}
 [\mathcal{O}^t_{3x}]^{\alpha_{1}} & [\mathcal{O}^t_{3y}]^{\alpha_{2}}  
 \d^c_{\alpha_{1}} \d^c_{\alpha_{2}}  W   =6 t^{-2} |\gamma^{t\mu}| P\left(\gamma^{t\mu};x-y\right)^3 \nn
& = \frac{6\mu^2}{(4\pi^2)^3}\frac{1}{\left(x^2+(t \mu)^{-2}\right)^3} \nn
&=\frac{6\mu^2}{(4\pi^2)^3}\Bigg[-\frac{1}{32}\dot \partial^4 \frac{\log \left[\left(x-y\right)^2\mu^2\right]}{\left(x-y\right)^2}+\frac{\pi^2}{8}\left( 1+2\log t \right) \partial^2 \delta(x-y)  \nn
&  \phantom{= \frac{6\mu^2}{(4\pi^2)^3} - } \mbox{} + \frac{\pi^2}{2}t^2 \mu^2\delta(x-y)  \Bigg]+o\left(t^0\right).
\end{align} 
The second term on the right-hand side of~\refeq{GR33} can be written as
\begin{align}
C^{t\,\alpha_1}_{3x3y} [\mathcal{O}^t_{\alpha_1}]^{\alpha_2} \partial^c_{\alpha_2} W  & =
- C^{t\,\alpha}_{3x3y} \langle  [\mathcal{O}^t_{\alpha}]\rangle \nn
& = - C^{t\,0z}_{3x3y}  \langle  [\mathcal{O}^t_{0z}]\rangle \nn 
& =- \left[\frac{3}{128 \pi^4}\left(\log t\right) + 2 E_2 \right] \mu^{2}\partial^2 \delta(x-y) -2 \mu^{4}t^{2}E_0\delta(x-y),
\label{hola}
\end{align}
where in the second equality we have used~\refeq{ExpValEig0} to discard all the directions $\alpha\neq 0$ in the sum, whereas in the third one we have used~\refeq{O0} and~\refeq{CTC0}.
Finally, taking the limit $t\to \infty$ we obtain
\begin{equation}
G^R_{3x3y}=-\frac{3\mu^2}{2^{10}\pi^6}\dot \partial^4 \frac{\log \left[\left(x-y\right)^2 \mu^2\right]}{\left(x-y\right)^2}+\mu^2\left( \frac{3}{2^8\pi^4}-2E_2 \right) \partial^2 \delta(x-y). \label{finalGR33}
\end{equation} 
The terms with $\log t$ have cancelled out and the result precisely agrees with~\refeq{reg33} for $E_2= \xi_3+3/(512\pi^4)$, as appropriate for the cutoff propagator we are using.

The two calculations of $G^R_{3x3y}$ that we have presented are based on the Wilsonian analysis of Section~\ref{sec:GausNormal}. In the last one, the exact RG flows have been used to find renormalised operators and counterterms that render the correlator finite. These objects can also be obtained without explicit Wilsonian information in the traditional way, just requiring that the UV divergences are cancelled in the correlation functions. Let us sketch the standard calculation to connect it with the one in this subsection. 
The starting point is the bare correlator
\beq
\left.\partial^c_{3x} \partial^c_{3y}\right|_{(0,\Lambda)}  W= \left|\gamma^{\Lambda}\right|\langle \omega(x)^3 \omega(y)^3 \rangle_{(0,\Lambda)} 
\label{CutoffFunc},
\eeq
which is given by the free-field diagrams $\mathcal{K}_1$ and $\mathcal{K}_2$ in Fig.~\ref{fig:S3S3}, with $\mu \to \Lambda= t\mu$. These diagrams are singular when $\Lambda \to \infty$. Both of them contain non-local divergences for separate points, $x \neq y$. The one in diagram $\mathcal{K}_1$ can be compensated by a multiplicative renormalisation of the operator $\omega^3$, while the ones in diagram $\mathcal{K}_2$ can be cancelled by adding to the action a counterterm proportional to $\omega$, which gives the contributions $\mathcal{K}_{3,4,5}$ in Fig.~\ref{fig:S3S3}. This linear renormalisation can be interpreted as the matrix renormalisation of the operator $\omega^3$ given by~\refeq{Ot1} and~\refeq{Ot3}. After it, only a local divergence for coincident points $x \sim y$ remains. The counterterm that cancels it is a local contribution to the vacuum energy, which can be identified with the non-linear contribution in~\refeq{hola}.

Of course, there is some freedom in the choice of counterterms that do the job. However, imposing the minimal subtraction  conditions written at the end of Section~\ref{sec:Correlation} we arrive at the same renormalised operators and counterterms given above. Therefore, we also obtain the same renormalised function $G^R_{3x3y}$ in~\refeq{finalGR33} and~\refeq{reg33}. This illustrates the general result, proven in Section~\ref{sec:Correlation}, that minimal subtraction leads to renormalised correlators that coincide with cutoff correlators in normal coordinates. In other words: in minimal subtraction schemes, the renormalised couplings can be understood as the couplings in a normal parametrisation of the action.  We also stress that the renormalised correlator $G^R_{3x3y}$ itself retains Wilsonian information about normal coordinates through the renormalisation scale. Indeed, because the renormalised correlators are equal to bare correlators at scale $\mu$ in normal coordinates, we know that they must obey Callan-Symanzik equations. A double functional differentiation of~\refeq{CallanSymanzik} with $\Lambda=\mu$ leads to
\beq
\mu \frac{\d}{\d \mu} G^R_{3x3y} = 
 2 \lambda_{(3)} G^R_{3x3y}  - 2 \beta^{0z}_{3x3y}(\gamma^\mu) G^R_{0z}. \label{CS33}
\eeq
Using~\refeq{finalGR33} in the left-hand side of~\refeq{CS33} we get
\beq
\mu \frac{\d}{\d \mu} G^R_{3x3y} = 2 G^R_{3x3y} + \frac{3 \mu^2}{128 \pi^4} \dot{\d}^2 \delta(x-y).
\eeq
Taking into account the obvious result for the renormalised vacuum energy, $G^R_{0z} = - \mu^4$, and comparing non-local and local pieces in~\refeq{CS33}, we find $\lambda_{(3)}=1$ (that is, $\Delta_{(3)}=3$) and $\beta^{0z}_{3x3y}(\gamma^\mu)$ as in~\refeq{beta0}.
We emphasize that this also holds for any regulator and in any renormalisation scheme consistent with minimal subtraction. This is particularly simple in mass-independent methods. For instance, we could simply compute the diagrams with the original unregularised propagator and use dimensional regularisation to make sense of the resulting expressions. Then, diagram $\mathcal{K}_2$ of Fig.~\ref{fig:S3S3} vanishes identically, while diagram $\mathcal{K}_2$ directly gives~\refeq{reg33} plus a pole in $1/(d-4)$, which is cancelled by a local counterterm, see~\cite{Dunne:1992ws, delAguila:1999bd}. The same result can be found even more directly, without explicit regularisation, in differential renormalisation~\cite{Freedman:1991tk}. It is remarkable that these mass-independent renormalisation schemes produce renormalised correlators associated to normal coordinates, which carry all the local information near the fixed point about the exact RG flows in these coordinates. The exact beta functions are equal to their Gell-Mann-Low counterparts, up to residual scheme dependence within minimal subtraction. The fact that only resonant terms appear can be understood by dimensional analysis in the absence of dimensionful parameters in the regularisation and renormalisation conditions.


\section{Conclusions}
\label{sec:Conclusions}

We have analysed the details of the relation between the exact RG, with local couplings, and the renormalisation of $n$-point correlation functions of composite operators (including irrelevant ones) at a fixed point. Our main tool has been the flexibility of choosing different, non-linearly related, coordinate systems. Among them, we have singled out {\em normal coordinates}, in which the RG flows take their simplest form, and {\em renormalisation coordinates}, which are scale dependent. Working with different parametrisations and making sense of them has been greatly facilitated by an intrinsic, geometric description of the RG flows and the renormalisation process. This formalism generalises the one in~\cite{Dolan:1994pq} and~\cite{Polonyi:2000fv}. In particular, the cutoff scale is incorporated in the description of theory space; it plays an active role not only in the RG flows but also in dealing with derivatives of local couplings, which are ubiquitous.   

We have provided explicit answers to the questions i) and ii) posed in the introduction. We have shown that the renormalised correlation functions are identical to correlation functions defined in the regularised theory at a finite cutoff, which is identified with the renormalisation scale. This result is given in~\refeq{tildeR}. 
The coordinates $\tilde{c}$ that define the cutoff correlators are simply related to the RG flows by~\refeq{rf} (with $c=\tilde{c})$. 

The connection between renormalisation and exact RG is specially transparent in minimal subtraction schemes, which give rise to bare correlators defined as functional derivatives in normal coordinates. Then, the exact beta functions are identified with the Gell-Mann-Low beta functions (and conformal anomalies) in minimal subtraction schemes. Therefore, the exact RG flows around the fixed point, in their simplest form, can be obtained from standard diagrammatic calculations of correlation functions of general composite operators. We find interesting the fact that calculations in mass-independent methods, such as dimensional regularisation with minimal subtraction, can provide all the perturbative information about the exact RG flows, which are defined with a dimensionful regulator. This suggests that it might be possible to define the exact RG in such a way that only normal coordinates of the Wilson actions are allowed. 
Aside from this speculation, our findings provide a precise relation between the exact RG near a fixed point and intrinsic properties of the CFT at the fixed point. Indeed, the singularities of the CFT determine the scaling of renormalised operators (given by conformal dimensions) and, to a large extent, the counterterms, which as we have just emphasized contain all the local information of the RG flows. The structure of these singularities is in turn fixed by basic properties of the theory, such as the Wilson coefficients of the OPE. It would be interesting to explore along these lines how the consistency conditions of the CFT used in the bootstrap program are implemented in the exact RG.

In the case of non-trivial UV fixed points, it may be convenient to perform the renormalisation procedure at a point on the critical surface different from the fixed point, when this critical point is under better control, for instance by perturbation in some small parameter. We have shown that the results highlighted above also hold in this case: the renormalisation at the critical point is related to the RG flows around the fixed point. This is pertinent in particular to computations on the gravity side of gauge/gravity duals. 
As shown in~\cite{Lizana:2015hqb}, the fixed points are quite involved in the Wilsonian holographic description. Therefore, it is easier to renormalise around a critical point given by vanishing Dirichlet boundary conditions. This was done in~\cite{Lizana:2015hqb} for correlators of relevant operators, using only Dirichlet conditions (as in standard holographic renormalisation~\cite{deHaro:2000vlm}), which are dual to single-trace deformations. However, to address the mixing under renormalisation of irrelevant single-trace operators with multi-trace ones, the corresponding renormalised correlators require a generalisation of the boundary conditions. In~\cite{LPVholography}, the same methods in this paper are used to systematically define the complete holographic renormalisation of arbitrary correlation functions and to find its relation with the Wilsonian RG near the fixed point. 

Finally, in this paper we have considered theories defined in flat Euclidean space. It would be interesting to generalise the formalism to curved spaces. In this case, the metric should be treated as a coupling, so the space-time geometry will be expected to evolve under the RG evolution. Such a generalisation might be useful, for example, to calculate conformal anomalies in strongly-interacting CFT using Wilsonian methods. It would also connect with holographic RG flows, in which the metric backreacts under deformations of the dual CFT.

\acknowledgments

It is a pleasure to thank Tim Morris for sharing with us his insight about the exact renormalisation group. This work has been supported by the Spanish  MICINN project FPA 2013-47836-C3-2-P, the MINECO project FPA2016-78220-C3-1-P and by the European Commission through the contract PITN-GA-2012-316704 (HIGGSTOOLS).

\bibliographystyle{JHEP}
\bibliography{correlators}

\end{document}